\documentclass[useAMS,usenatbib]{mn2e}
\usepackage{epsfig}
\usepackage{graphicx}
\usepackage{journal_names}
\usepackage[english]{babel}

\usepackage{amsmath}
\usepackage{amssymb}
\usepackage{amsthm}
\usepackage{fancyhdr,txfonts} 

\usepackage{natbib}
\usepackage{sidecap}
\usepackage{subfig}

\usepackage{color}
\usepackage{wasysym}
\usepackage{url}

\usepackage[para]{threeparttablex}

\usepackage[multiple]{footmisc}

\makeatletter
\newcommand\footnoteref[1]{\protected@xdef\@thefnmark{\ref{#1}}\@footnotemark}
\makeatother

\title[The Build-up of The Red Sequence at $z\sim1$]{The Accelerated Build-up of the Red Sequence in High Redshift Galaxy Clusters}
\author[P. Cerulo et al.]{\parbox{\textwidth}{P. Cerulo$^{1}$\thanks{E-mail:
pcerulo@astro.swin.edu.au}, W. J. Couch{$^{1,2}$}, C. Lidman$^{2}$, R. Demarco$^{3}$, M. Huertas-Company$^{4}$, S. Mei$^{4}$, R. S\'{a}nchez-Janssen$^{5}$, L. F. Barrientos$^{6,7}$, R. P. Mu\~{n}oz$^{6}$}\vspace{0.4cm}\\
\parbox{\textwidth}{$^{1}$Centre for Astrophysics and Supercomputing, Swinburne University of Technology, PO Box 218, Hawthorn, VIC 3122, Australia\\
$^{2}$Australian Astronomical Observatory, PO Box 915, North Ryde, NSW 1670, Australia \\
$^{3}$Department of Astronomy, Universidad de Concepcion, Casilla 160-C, Concepcion, Chile\\
$^{4}$GEPI, Paris Observatory, 77 av. Denfert Rochereau, 75014 Paris, France\\
$^{5}$NRC Herzberg Astronomy and Astrophysics, 5071 West Saanich Road, Victoria, BC, V9E 2E7, Canada \\
$^{6}$Instituto de Astrofisica, Pontificia Universidad Catolica de Chile\\
$^{7}$Millennium Institute of Astrophysics, Chile\\}}

\begin{document}


\pagerange{\pageref{firstpage}--\pageref{lastpage}} \pubyear{2012}

\maketitle

\label{firstpage}

\begin{abstract}
We analyse the evolution of the red sequence in a sample of galaxy clusters at redshifts $0.8<z<1.5$ taken from the HAWK-I Cluster Survey (HCS). The comparison with the low-redshift ($0.04<z<0.08$) sample of the WIde-field Nearby Galaxy-cluster Survey (WINGS) and other literature results shows that the slope and intrinsic scatter of the cluster red sequence have undergone little evolution since $z=1.5$. We find that the luminous-to-faint ratio and the slope of the faint end of the luminosity distribution of the HCS red sequence are consistent with those measured in WINGS, implying that there is no deficit of red galaxies at magnitudes fainter than $M_V^*$ at high redshifts. \textcolor{black}{We find that the most massive HCS clusters host a population of bright red sequence galaxies at $M_V < -22.0$ mag, which are not observed in low-mass clusters. Interestingly, we also note the presence of a population of very bright ($M_V < -23.0$ mag) and massive (${\log(M_*/M_\odot)} > 11.5$) red sequence galaxies in the WINGS clusters, which do not include only the brightest cluster galaxies and which are not present in the HCS clusters, suggesting that they formed at epochs later than $z=0.8$. The comparison with the luminosity distribution of a sample of passive red sequence galaxies drawn from the COSMOS/UltraVISTA field in the photometric redshift range $0.8<z_{phot}<1.5$ shows that the red sequence in clusters is more developed at the faint end, suggesting that halo mass plays an important role in setting the time-scales for the build-up of the red sequence.}   
\end{abstract}


\section{Introduction}

The evolution of galaxies is driven by a combination of competing internal and external mechanisms. The first are related to galaxy stellar mass, while the latter are related to the environment in which galaxies reside. It has been demonstrated that stellar mass and environment both conspire in quenching star formation (e.g.\ \citealt{Kauffmann_2004}, \citealt{Baldry_2006}, \citealt{Peng_2010}), while a wealth of environmental mechanisms have been shown in theoretical studies to promote or attenuate star formation. Processes such as galaxy-galaxy mergers \citep{Lavery_1988}, harassment and tidal interactions (\citealt{Moore_1998}, \citealt{Bekki_2011}), strangulation \citep{Larson_1980}, and ram pressure stripping \citep{Gunn_1972} are all likely to take place in the dense environments of clusters and groups of galaxies. Although it is not yet clear which of these mechanisms is the \textcolor{black}{main environmental driver of galaxy evolution, in a recent paper, \cite{Peng_2015} argue that in the local universe most galaxies were quenched over long ($\sim 4$ Gyr) timescales (\itshape{strangulation}}). These conclusions are in agreement with the predictions of the theoretical works of \cite{Taranu_2014} and \cite{Bahe_2015} for clusters of galaxies.

Galaxy clusters are the most massive virialised systems in the Universe and, with their variety of environments, ranging from the dense cores to the sparse outskirts, provide natural laboratories for the study of the environmental drivers of galaxy evolution. One of the principal features in galaxy clusters is the tight and prominent red sequence in the colour-magnitude diagram, which can be observed up to redshifts $z \sim 2$ (\citealt{Tanaka_2010}, \citealt{Gobat_2011}, \citealt{Spitler_2012}, \citealt{Stanford_2012}, \citealt{Andreon_2014}). 


The colour-magnitude relation along the red sequence can be modelled by a straight line, and \cite{Kodama_1997} showed that its physical interpretation is that of a mass vs metallicity relationship with the intrinsic scatter associated with galaxy stellar age \citep{Bower_1992, Jaffe_2011}. The investigation of the build-up of the red sequence is a powerful tool to understand the mechanisms responsible for the shut-down of star formation. So far two aspects have been considered, the evolution of the red sequence slope, zero-point and scatter, and the study of the build-up as a function of stellar mass. 

Observations of clusters up to redshift $z=1.8$ have shown that there has been little evolution in the slope of the red sequence, which at all redshifts is found to be negative, \textcolor{black}{suggesting that the main features of the red sequence were already established at those epochs (\citealt{Ellis_1997}, \citealt{Stanford_1998}, \citealt{Lidman_2004}, \citealt{Lidman_2008}, \citealt{Ascaso_2008}, \citealt{Mei_2009}, \cite{Stott_2009}, \citealt{Papovich_2010}, \citealt{Snyder_2012}).} Yet theoretical works based on the hierarchical merging paradigm have not always been successful in predicting slopes that are consistent with those observed in distant clusters. For example, the hydrodynamical and N-body simulations of \cite{Romeo_2008} predict a strong evolution in the slope of the cluster red sequence, which at $z=0.8$ flattens and then turns positive, while the semi-analytical models of \cite{Menci_2008} predict a non-evolving flat red sequence. This highlights a major deficiency in these models. Recently, \cite{Merson_2015_arxiv}, using the semi-analytical model {\ttfamily{GALFORM}}, have shown that one can successfully reproduce red sequences with negative slopes, consistent with observations of clusters at $0.8<z<1.5$. However, the authors underline that, with their prescriptions, the cluster luminosity function is underestimated at $L^*$. The adjustment of some of the parameters, such as the dust obscuration law or AGN and supernova feedback efficiencies, results in model luminosity functions in better agreement with the observations but in red sequences that deviate significantly from the observations at bright magnitudes. In an upgraded version of their models, \cite{Romeo_2015} were able to reproduce a milder evolution of the red sequence slope, which remains negative up to redshifts $z = 1.5$. Generally speaking, the correct reproduction of the red sequence requires some post-processing of the outputs of semi-analytical models (see e.g.\ \citealt{Ascaso_2015}).

The hydrodynamical simulations of \cite{Gabor_2012}, which implement an empirical quenching mechanism based on the regulation of gas cooling and inflow towards the galaxy, are able to reproduce, at $z=1$, red sequences with negative slopes and a deficit of galaxies at stellar masses $M_* < 10^{10.5} \mbox{ } M_\odot$. Such a trend is in agreement with observations of galaxy clusters at $z<1.6$ that find that the faint end of the red sequence becomes gradually less populated at higher redshifts (e.g. \citealt{Capozzi_2010}, \citealt{Bildfell_2012}, \citealt{De_Lucia_2007}, \citealt{Gilbank_2008}, \citealt{Huertas_2009b}, \citealt{Lemaux_2012}, \citealt{Rudnick_2012}, \citealt{Fassbender_2014}). \textcolor{black}{These observations suggest that low-mass galaxies formed their stars later and joined the red sequence at lower redshifts, in agreement with \cite{Demarco_2010}, \cite{Muzzin_2012} and \cite{Nantais_2013b}, who showed that high-mass galaxies in clusters at $0.8 < z < 1.3$ are on average older and less star-forming than their lower-mass counterparts.}

This scenario, which was first proposed in \cite{Tinsley_1968}, is in agreement with observations of galaxies in the field and in low-density environments (e.g,: \citealt{Cowie_1996}, \citealt{Pozzetti_2010}), but is still a matter of debate in clusters, where the environment is expected to play a major role in quenching star formation in low-mass satellite galaxies. Thus, in the works of \cite{Andreon_2008}, \cite{Andreon_2014}, \cite{Crawford_2009}, \cite{Lidman_2008}, and \cite{De_Propris_2013}, no deficit of galaxies is found at the faint end of the red sequence in clusters at $z<1.8$ (see also \citealt{Brown_2008} for similar conclusions in the field). These authors, however, do not exclude that the deficit may be a property of some clusters.

An interesting perspective is offered by the results of \cite{Lemaux_2012}, who dissected the Cl1604 supercluster at $z \sim 0.9$ in its constituent groups and clusters, finding that the least virialised and massive structures in the system presented a deficit of galaxies at bright and faint luminosities. This result suggests that cluster halo mass may play an important role in setting the timescales for the build-up of the red sequence (see also \citealt{Tanaka_2010} and \citealt{Tanaka_2013}). As shown by \cite{Gabor_2015}, the most massive haloes also host the densest environments and, therefore, investigating clusters with different halo masses may be another way of characterising the environments of individual clusters.


In this paper we present a comprehensive analysis of the properties of the red sequence in a sample of 9 galaxy clusters at $0.8<z<1.5$ in the HAWK-I Cluster Survey (HCS, \citealt{Lidman_2013}) for which optical and infrared (IR) imaging data from space-and ground-based observations, as well as spectra, are available. We have already presented a study of the properties of the red sequence in one of the HCS clusters, namely  XMMU J1229+0151, at $z=0.98$, in \cite{Cerulo_2014}, where we also discussed the analysis method developed for the study of clusters in the HCS. \textcolor{black}{This is the first of two papers in which we apply the method of \cite{Cerulo_2014} to the entire HCS sample. The second paper of this series will be focused on the study of galaxy morphology and is currently in preparation (Cerulo et al. 2015b). The aim of the present paper is to study the evolution of the parameters of the red sequence and the build-up of the red sequence as a function of galaxy luminosity.} The paper is organised as follows. We describe the observations and data reduction in Section 2, while Section 3 discusses the photometry and the estimation of cluster membership. The results of the measurements of the red sequence parameters, luminous-to-faint ratio, and luminosity distribution are presented in Section 4 and discussed in Section 5. Section 6 summarises our results and draws the conclusions of the analysis.

Throughout the paper we adopt a $\Lambda CDM$ cosmology with $\Omega_\Lambda = 0.73$, $\Omega_{m} = 0.27$, and $H_0 = 71.0$ km $\cdot$ s$^{-1}$ $\cdot$ Mpc$^{-1}$. Unless otherwise stated, all magnitudes are quoted in the AB system \citep{Oke_1974}. We define $R_{200}$ as the physical radius, measured in Mpc, including the region where the total matter density (baryonic and non-baryonic) is 200 times higher than the critical density at the redshift of each cluster.

\section{Observations and data reduction} 

\subsection{The HAWK-I Cluster Survey}

The HAWK-I Cluster Survey (HCS, PI: Lidman) is a near infrared (NIR) observing programme carried out with the High Acuity Wide-field K-band Imager (HAWK-I, \citealt{Pirard_2004}) on the European Southern Observatory (ESO) 8.2 m Very Large Telescope (VLT) with the aim of studying galaxy populations in clusters at redshifts $z > 0.8$. The HCS sample currently consists of 9 clusters, seven taken from the Hubble Space Telescope (HST) Cluster Supernova Survey \citep{Dawson_2009}, the cluster RXJ0152.7-135 (RX0152), which is part of the Advanced Camera for Surveys Intermediate Redshift Cluster Survey (\citealt{Ford_2004}, \citealt{Postman_2005}, \citealt{Mei_2009}), and one cluster from the Spitzer Adaptation of the Red Sequence Cluster Survey (SpARCS J003550-431224, $z=1.34$, \citealt{Muzzin_2012}, \citealt{Lidman_2012}). The sample is composed of a mixture of optically, IR, and X-ray detected clusters, thus ensuring that a broad range of cluster morphologies, from spiral-rich to cD-dominated, are considered (see \citealt{Bahcall_1977} for a review on galaxy cluster morphology).

All the clusters were observed at least in the HAWK-I Ks band, while only clusters at $z>1.1$ were observed in both the J and Ks bands. The observations and data reduction of these images are discussed in \cite{Lidman_2013}. For all the clusters, the HAWK-I images cover a $\sim 10' \times 10'$ field of view with a final image quality, parametrised by the Full Width at Half Maximum (FWHM) of the Point Spread Function (PSF), in the range $FWHM = 0.3''-0.4''$ in both the J and Ks bands. 

The cluster RDCS J1252.9-2927 (RDCS1252), observed in the Js and Ks bands of the Infrared Spectrometer And Array Camera (ISAAC, \citealt{Moorwood_1998b}), previously mounted on the ESO/VLT and now decommissioned, was also added to the HCS sample and studied together with the other clusters. This brings the final sample to a total of 10 clusters. The NIR observations and data reduction for RDCS1252 are discussed in \cite{Lidman_2004}, while we summarise the main properties in Table \ref{table2}. The final mosaicked images, in both the Js and Ks bands, have $4.7' \times 4.7'$ fields of view with image qualities $FWHM \sim 0.4''$.


In this work and in Cerulo et al. (2015b) we will study all the HCS clusters except SpARCS J003550-431224 for which no HST Advanced Camera for Surveys (ACS) data are currently available. Table \ref{table1} shows the global properties of the clusters, while Table \ref{table2} summarises the optical and NIR observations, which are outlined in the following sections..


\begin{table*}
  \caption{The HAWK-I Cluster Survey (HCS) sample with the clusters listed in order of increasing redshift. The dark matter halo masses $M_{DM}$ in the fifth column from the left are taken from \protect\cite{Jee_2011}. These masses are all estimated with a weak lensing analysis carried out on the ACS images.}
  \begin{minipage}{14 cm}
  \begin{tabular}{|l|l|l|c|c|c|}
    \hline
     \multicolumn{1}{|c}{Cluster Name}  & \multicolumn{1}{c}{$\alpha$ (J2000)} & \multicolumn{1}{c}{$\delta$ (J2000)} & \multicolumn{1}{c}{Redshift} &  \multicolumn{1}{c}{$M_{DM}$} & \multicolumn{1}{c|}{Spectroscopically} \\
     \multicolumn{1}{|c}{}  & \multicolumn{1}{c}{} & \multicolumn{1}{c}{} & \multicolumn{1}{c}{} & \multicolumn{1}{c}{($10^{14} M_\odot$)} & \multicolumn{1}{c|}{Confirmed}   \\
     \multicolumn{1}{|c}{} & \multicolumn{1}{c}{} & \multicolumn{1}{c}{} & \multicolumn{1}{c}{} & \multicolumn{1}{c}{} & \multicolumn{1}{c|}{Members}  \\                  
    \hline
     \hline
     RX J0152.7-135 (RX0152) & 01:53:00 & -13:57:00 & 0.84 & $4.4^{+0.7}_{{-}{0.5}}$ & 134 \\
     RCS 2319.8+0038 (RCS2319) & 23:19:53.9 & +00:38:13 & 0.91 & $5.8^{+2.3}_{{-}1.6}$ & 58  \\
     XMM J1229+0151 (XMM1229) & 12:29:28.8 & +01:51:34 & 0.98 & $5.3^{+1.7}_{{-}1.2}$  & 18   \\
     RCS 0220.9-0333 (RCS0220) & 02:20:55.7 & -03:33:19 & 1.03 & $4.8^{+1.8}_{{-}1.3}$ &  21    \\
     RCS 2345-3633 (RCS2345) & 23:45:27.3 & -36:32:50 & 1.04 & $2.4^{+1.1}_{{-}0.7}$  &  29   \\
     XMM J0223-0436 (XMMU0223) & 02:23:03.7 & -04:36:18 & 1.22 & $7.4^{+2.5}_{{-}1.8}$  & 20    \\
     RDCS J1252.9-2927 (RDCS1252) & 12:52:00 & -29:27:00 & 1.24 & $6.8^{+1.2}_{{-}1.0}$ &  42   \\
     XMMU J2235.3-2557 (XMMU2235) & 22:35:00 & -25:57:00 & 1.39 & $7.3^{+1.7}_{{-}1.4}$  & 25   \\
     XMM J2215-1738 (XMMXCS2215) & 22:15:58.5 & -17:38:02 & 1.45 & $4.3^{+3.0}_{{-}1.7}$ &  26  \\
  \hline
  \end{tabular}
\end{minipage}
\label{table1}
\end{table*}

\subsection{Advanced Camera for Surveys (ACS)}

The ACS data for 8 of the HCS clusters are taken from the sample of the HST Cluster Supernova Survey (\citealt{Dawson_2009}), which consists of deep images taken in the F775W ($i_{775}$) and F850LP ($z_{850}$) bands collected over multiple HST visits on each cluster. In particular, each visit consisted of 3 or 4 exposures in the $z_{850}$ band and at least 1 in the $i_{775}$ band. This resulted into an average exposure time of 3,000 s in the $i_{775}$ band and 10,000 s in the $z_{850}$ band for all the clusters. The images were processed with the standard Space Telescope Science Institute (STScI) data reduction pipeline with the most up to date calibration files and were combined using {\ttfamily{Multidrizzle}} \citep{Fruchter_2002, Koekemoer_2002} to the final pixel scale 0.05$''$/pixel. The field covered by these images is on average $5' \times 5'$ with a resulting image quality $FWHM \sim 0.09''$ in both bands.

The $i_{775}$- and $z_{850}$-band images for the cluster RX0152 were taken over multiple orbits of HST. Four pointings, arranged in a 2 by 2 pattern, were used to increase the depth of the final image within 1$'$ of the cluster centre. This resulted in deeper exposures with respect to the average of the HST Cluster Supernova sample (see Table \ref{table2}). In addition to the F775W and F850LP filters, RX0152 was also observed in the F625W ($r_{625}$) filter following the same strategy of the $i_{775}$ and $z_{850}$ observations. The RX0152 images were reduced and combined with the APSIS pipeline \citep{Blakeslee_2003b} to a final pixel scale of 0.035$''$/pixel. The total field covered by the images is $6' \times 6'$ in all the three bands with PSF FWHM $\sim 0.07''$.

\subsection{Wide Field Camera 3 (WFC3)}

The clusters XMMU J1229+0151 (XMM1229), RDCS J1252.9-2927 (RDCS1252), and XMMU J2235.3-2557 (XMMU2235) were also observed in the F105W (Y), F110W, F125W (J), and F160W (H) bands of the IR channel of the HST Wide Field Camera 3 (WFC3) during Program 12051 (P. I. S. Perlmutter), aimed at the calibration of the sensitivities of the HST Near Infrared Camera and Multi-Object Spectrometer (NICMOS) and WFC3 for faint objects \citep{Rubin_2015}. The XMM1229 observations and data reduction are presented in \cite{Cerulo_2014}. The same procedure for image co-addition and combination was adopted for RDCS1252 and XMMU2235, although for these two clusters we used the latest version of the {\ttfamily{DrizzlePac}}\footnote{The latest version of the {\ttfamily{DrizzlePac}} can be downloaded from: \url{http://www.stsci.edu/hst/HST\_overview/drizzlepac}}, released by STScI in June 2012, after the reduction of the XMM1229 data. This package contains an updated and revised implementation of the {\ttfamily{Multidrizzle}} algorithm.

\textcolor{black}{The combined WFC3 images of the HCS clusters, drizzled to the final pixel scale $0.06''$/pixel, cover an area of $3' \times 3'$, The observed image quality, resulting from the convolution between the PSF of the instrument and the pixel response function of the IR detector, is $(FWHM)_{obs} \sim 0.2''$ in all images. After subtracting in quadrature the width of the pixel response function ($0.128''$), we find that the intrinsic image quality is in the range $0.12'' < (FWHM)_{int} < 0.18''$ for all the three clusters observed with WFC3.}

\subsection{Infrared Spectrometer And Array Camera (ISAAC)}

The ISAAC observations of the 3 clusters RCS 2319.8+0038 (RCS2319, $z=0.91$), RCS 0220.9-0333 (RCS0220, $z=1.03$), and RCS 2345-3633 (RCS2345, $z=1.04$) (R.\ P.\ Mu\~{n}oz's PhD thesis 2009) are part of a NIR observing programme aimed at the study of the build-up of the red sequence in a sub-sample of clusters of the Red Sequence Cluster Survey (RCS, \citealt{Gladders_2005}) . The programme was distributed over four observing runs, which took place between 2001 and 2003 and targeted 15 clusters at $z\sim1$. The three clusters included in the HCS sample were observed during the ESO programmes 70.A-0378 and 71.A-0345 (P.I.\ L.\ F. Barrientos). 

ISAAC \citep{Moorwood_1998b} was an IR imager and spectrograph optimised for observations in the range $1 \mu m < \lambda < 5 \mu m$, previously mounted on the ESO/VLT and decommissioned in 2013. The three clusters were all observed with the short-wavelength arm, equipped with a 1024 $\times$ 1024 HgCdTe Hawaii detector with pixel size $0.1484''$. RCS0220 was observed with the J filter, while RCS2319 and RCS2345 were both observed with the Js filter. The wavelength range covered by the J filter is broader than that covered by Js. However, we find that the observed AB $(J-Js)$ colour at $0.90 < z < 1.05$ predicted for a model spectral energy distribution (SED) taken from the \cite{Bruzual_2003} library, with formation redshift $z_f=4.75$, \cite{Salpeter_1955} initial mass function (IMF), exponentially declining star-formation history with $\tau=1$ Gyr, and solar metallicity, is $(J-Js) = 0.001$ mag. \textcolor{black}{With such a small colour term, we assume $J=Js$ throughout the rest of the paper\footnote{The model SEDs used in this paper were built with the EzGaL Python package \citep{Mancone_2012_ezgal}, which can be downloaded at \url{http://www.baryons.org/ezgal/}}.}

RCS0220 was observed for a total of 45 minutes, while both RCS2319 and RCS2345 were observed for 54 minutes. The observations consisted in a series of dithered exposures of 30 s each with the telescope randomly offset within a square region 38$''$ wide. Each exposure was dark and sky subtracted, and was finally flat-field corrected. The photometric calibration was performed on stars of the NICMOS \citep{Persson_1998} and UKIRT\footnote{The  United Kingdom Infrared Telescope} \citep{Hawarden_2001} photometric standard star catalogues which were observed at low airmasses during the same observing nights in which the science data were taken. Once the magnitude zero-point for each cluster was estimated, all the images were re-calibrated to a common magnitude zero-point $J_{Vega} = 28.0$ mag.

The ISAAC observations of the HCS clusters are summarised in Table \ref{table2}.

\subsection{Spectroscopy}

\textcolor{black}{The HCS clusters have been targeted in various spectroscopic follow-up programmes of the HST Cluster Supernova Survey conducted at the Keck and VLT telescopes. In this work we use all the redshifts obtained from those observations. The cluster RCS2319, which belongs to the RCS2319+00 supercluster, was also part of an extensive survey of the supercluster conducted at the Magellan, VLT, Subaru, and Gemini-North telescopes. This data-set is discussed in \cite{Faloon_2013}, and we refer to that paper for a description of the observations and data reduction. We included the redshift catalogue from the RCS2319+00 data-set in the HCS sample.}

\textcolor{black}{The cluster XMMU J0223-0436 (XMMU0223, $z=1.22$) falls in the field of view of the VIMOS Public Extragalactic Redshift Survey (VIPERS, \citealt{Garilli_2014, Guzzo_2014}), and the redshifts coming from this sample were added to the HCS sample.}

\textcolor{black}{The cluster RX0152 was observed with FORS1 and FORS2\footnote{FOcal Reducer and low dispersion Spectrograph} at the ESO/VLT between 2005 and 2009  (see \citealt{Demarco_2010} for a summary of these observations), while RDCS1252 was observed in 2003, 2011, and 2012 with FORS2 (\citealt{Demarco_2007}, \citealt{Nantais_2013b}). In the present work we include all the redshifts available for RX0152 and RDCS1252 and used in the works of \cite{Demarco_2010} and \cite{Nantais_2013b}.}

\textcolor{black}{The clusters XMM1229, RCS2319, and RCS0220 were the targets of deep spectroscopic observations that were conducted at the Keck and Gemini North telescopes, with the aim of acquiring high signal-to-noise ($S/N$) spectra to study stellar populations and increase the cluster spectroscopic sampling towards faint magnitudes ($z_{850} = 24.0$ mag). RCS2319 was observed with the Low-Resolution Imager Spectrograph (LRIS, \citealt{Oke_1995}) at the Keck I telescope, while XMM1229 and RCS0220 were observed with the Gemini North Multi-Object Spectrograph (GMOS-N, \citealt{Hook_2004_GMOS}). The redshifts of the galaxies targeted in these observations were measured with the RUNZ software{\footnote{RUNZ can be downloaded from: \url{http://www.physics.usyd.edu.au/~scroom/runz/}}}, which is based on the cross-correlation method of \citep{Tonry_1979}. The observations, data reduction, and redshift measurement will be presented in a forthcoming paper together with the analysis of the stellar populations in these and other HCS clusters.}

\begin{table*}
   \begin{threeparttable}
  \caption{Summary of the HCS imaging observations. The 90\% magnitude completeness limit is estimated as discussed in \protect\cite{Cerulo_2014} by inserting random simulated galaxies in the images and computing the fraction of recovered sources. The image quality FWHM quoted for WFC3 corresponds to the intrinsic PSF FWHM obtained by de-convolving the observed PSF by the detector pixel response function. Each line refers to an instrument; e.g.\ the third line of the XMM1229 entry refers to the WFC3 observations of this cluster.}
 \begin{tabular}{|c|c|c|c|c|} 
  \hline
      \multicolumn{1}{|c}{Cluster} & \multicolumn{1}{c}{Filter(Exposure Time)} & \multicolumn{1}{c}{Image Quality} & \multicolumn{1}{c|}{90\% magnitude} \\
       \multicolumn{1}{|c}{} & \multicolumn{1}{c}{($10^3$ s)} & \multicolumn{1}{c}{(FWHM) ('')} & \multicolumn{1}{c|}{completeness limit (mag)} \\                                                     
  \hline
  \hline
  RX0152         & $r_{625}$\tnote{b} (19.0), $i_{775}$\tnote{b} (19.2), $z_{850}$\tnote{b} (19.0)    & 0.07, 0.07, 0.08          & 26.7, 26.2, 25.7            \\
                 &  Ks (9.6)\tnote{d}                                             & 0.34                      & 23.4                        \\
 RCS2319       & $i_{775}$\tnote{b} (2.4), $z_{850}$\tnote{b} (6.8)                       & 0.10, 0.10,              & 26.8, 26.2                   \\
                & $Js_{ISAAC}$ (3.2)                                       & 0.58                      &  22.3                        \\
                 & Ks(9.6)\tnote{d}                                               &  0.39                    &  24.9                         \\
XMM1229      & R(1.14)\tnote{a}                                            & 0.63                     & 25.3                      \\
               & $i_{775}$\tnote{b} (4.2), $z_{850}$\tnote{b} (10.9)                      & 0.08, 0.09                & 25.0, 25.0                 \\
              & F105W\tnote{c} (1.3), F110W\tnote{c} (1.1), F125W\tnote{c} (1.2), F160W\tnote{c} (1.1)    & 0.11, 0.11, 0.13, 0.14    & 23.0, 23.2, 23.3, 23.5     \\
                & $J_{SofI}$ (2.3)                                       & 0.94                      & 22.4                       \\
                & Ks (11.3)\tnote{d}                                             & 0.34                      & 24.6                       \\
RCS0220        & $i_{775}$\tnote{b} (3.0), $z_{850}$\tnote{b} (14.4)                       & 0.10, 0.10               & 26.2, 26.7                    \\
                & $J_{ISAAC}$ (2.7)                                       & 0.45                     &   22.4                        \\
               & Ks (9.6)\tnote{d}                                              & 0.31                      &   24.2                        \\
 RCS2345       & $i_{775}$\tnote{b} (4.5), $z_{850}$\tnote{b} (9.7)                       & 0.094, 0.10                & 26.6, 26.4                    \\
                & $Js_{ISAAC}$ (3.2)                                     & 0.57                       & 21.9                           \\
                & Ks (9.6)\tnote{d}                                             & 0.31                       &  24.2                           \\
 XMMU0223         & $i_{775}$\tnote{b} (3.4), $z_{850}$\tnote{b} (14.02)                    & 0.10, 0.10,           & 25.5, 25.3                          \\
                 & J\tnote{d} (11.04), Ks\tnote{d} (9.6)                                 & 0.36, 0.34                &    24.9 , 23.44                     \\
 RDCS1252        & $i_{775}$\tnote{b} (29.9), $z_{850}$\tnote{b} (57.1)                    & 0.091, 0.097,              &   27.3, 26.1,                    \\
                 & F105W\tnote{c} (1.2), F110W\tnote{c} (1.1), F125W\tnote{c} (1.2), F160W\tnote{c} (1.2)   & 0.13, 0.15, 0.14, 0.14     &  26.1, 26.6, 25.9, 25.5          \\
                &    $J_{ISAAC}$ (86.6), $Ks_{ISAAC}$ (82.0)              &  0.43, 0.38              &      24.6, 24.2                  \\
XMMU2235        & $i_{775}$\tnote{b} (8.2), $z_{850}$\tnote{b} (14.4)                        & 0.094, 0.10              &    26.1, 26.2                     \\
                & F105W\tnote{c} (1.2), F110W\tnote{c} (1.1), F125W\tnote{c} (1.2), F160W\tnote{c} (1.2)     & 0.11, 0.12, 0.11, 0.14    &  25.04, 26.2, 25.8, 25.0          \\
                &               J\tnote{d} (10.6), Ks\tnote{d} (10.7)                     &   0.48, 0.36              &     22.6, 22.9                     \\
 XMMXCS2215    & $i_{775}$\tnote{b} (3.3), $z_{850}$\tnote{b} (16.9)                       & 0.093, 0.097,             &  24.4, 24.8,                      \\
                  &        J\tnote{d} (14.4), Ks\tnote{d} (9.6)                              &   0.47, 0.36              &   24.1, 24.5                      \\
\hline
\end{tabular}
\footnoterule
\begin{tablenotes}
    \item [a] FORS2 
    \item [b] ACS 
    \item [c] WFC3
    \item [d] HAWK-I
\end{tablenotes}
\label{table2}
 \end{threeparttable}
\end{table*}

\subsection{The Low Redshift Cluster Sample}

We use the spectroscopic sample of the WIde-field Nearby Galaxy-cluster Survey (WINGS, \citealt{Fasano_2006}) to compare the HCS red sequence with that of low-redshift clusters. WINGS is composed of 78 clusters in the redshift range $0.04 < z < 0.08$ observed in up to 5 photometric bands (U, B, V, J, K). A subsample of this survey, composed of 48 clusters, was followed up spectroscopically with the William Herschel Telescope (WHT) at the La Palma Observatory (Spain) and with the Anglo-Australian Telescope (AAT) at the Siding Spring Observatory (Australia). The sample contains 6120 galaxies in total with 3641 spectroscopically confirmed cluster members. The spectroscopic observations of the WINGS clusters are presented in \cite{Cava_2009}. Catalogues are publicly available and information for the download is given in \cite{Moretti_2014}. In the present work we use the B- and V-band photometric catalogues together with all the available redshifts. 

\textcolor{black}{WINGS targets clusters selected in the ROSAT All Sky Survey (\citealt{Ebeling_1996}, \citealt{Ebeling_1998}, \citealt{Ebeling_2000}), while in the HCS sample 3 clusters (RCS2319, RCS0220, and RCS2345) were optically detected and 6 clusters were detected in the X-rays. This difference in the sample selection techniques may lead to the construction of two intrinsically different samples and to a comparison between high- and low-redshift clusters which is significantly affected by systematics.}

\textcolor{black}{The X-ray detection of clusters, based on the measurement of the diffuse emission of the hot intracluster gas, does not privilege a particular cluster morphology or richness against the other. On the other hand, optical detection techniques based on the red sequence detection, such as that adopted in the RCS survey, are more sensitive to high cluster richness and high halo masses.}

\textcolor{black}{However, most of the WINGS sample overlaps with the Abell catalogues (\citealt{Abell_1958}, \citealt{Abell_1989}), which comprise rich systems detected on optical images. On the other hand, the X-ray emission from the intracluster gas becomes gradually fainter at high redshifts, allowing only the most massive distant clusters to be detected in X-ray surveys. Therefore, both the WINGS and HCS samples can be considered biased towards massive and rich systems and, consequently, the comparison between these two datasets should not be affected by large systematics. Interestingly, the cluster RCS2345 is a spiral-rich system with no prominent giant elliptical in its centre (see \citealt{Lidman_2013}). It is also the least massive cluster in the HCS sample, which demonstrates that optical samples of clusters may also host systems which are far from being rich and massive.}

We estimated the dark matter halo mass $M_{200}$ within $1 \times R_{200}$ of the centre of each WINGS cluster using the velocity dispersions provided in \cite{Cava_2009}. Under the assumption that each cluster is virialised, its halo mass can be approximated by the equation (\citealt{Finn_2005}, \citealt{Poggianti_2006}):
\begin{multline}
M_{200} = 1.2*10^{15} \left( \frac{\sigma}{1000 \mbox{ } km \cdot s^{-1}} \right)^3 \\
                   \times \frac{1}{\sqrt{\Omega_{\Lambda} + \Omega_m (1+z)^3}} h^{-1} \mbox{ } M_\odot
\end{multline}
where $\sigma$ is the cluster velocity dispersion.

\textcolor{black}{In order to investigate the systematics inherent in different mass measurement methods, we have estimated the masses of each HCS cluster with Equation 1 and compared with the available weak lensing estimates from \cite{Jee_2011}. Velocity dispersions were obtained from the available redshifts following the approach of \cite{Harrison_1974} and adopting a 3$\sigma$ clipping algorithm to remove field interlopers (see \citealt{Yahil_Davis_1977}). Estimates of the velocity dispersion exist for all the HCS clusters except RCS0220 and are summarised in Table 2 of \cite{Jee_2011}. Our estimates are all consistent with the literature results to within 2$\sigma$, although on average 1.3 times larger. This bias may be partly due to the fact that we do not correct for the additional broadening due to redshift uncertainties \citep{Danese_1980} because redshift errors are not available for all the galaxies in HCS.}

\textcolor{black}{We find that the mass estimates obtained in this way are on average 3 times larger than the weak lensing masses, although the measurements are still consistent (except in the cases of RX0152 and RDCS1252) within 3$\sigma$ with the weak lensing estimates. We also find that our halo mass estimates are on average 3 times larger than the masses obtained with Equation 1 using the velocity dispersions from the literature. These differences surely reflect the underlying overestimation of the velocity dispersion mentioned above. However, Equation 1 relies on the assumption that clusters are virialised, which may not apply to the entire HCS sample. As indeed shown in \cite{Sereno_2013}, the values of the dark matter concentration parameter in the most massive HCS clusters suggest that these systems have recently experienced mergers or the accretion of smaller haloes (groups).}

\textcolor{black}{As shown in \cite{Ramella_2007}, substructures in the galaxy distribution have been detected in 70\% of the WINGS clusters, suggesting that also these systems may have recently experienced the accretion of galaxy groups. Equation 1 may therefore constitute a simplification of a more complex dynamical picture for both the high- and low-redshift samples. However, for the purposes of this work, in which dynamical and dark matter properties of clusters are not investigated, Equation 1 is sufficient to obtain at least an approximate estimate of the halo mass in WINGS. In the remaining of the paper we will use the estimates of the halo mass from Equation 1 for WINGS and the weak lensing masses from \cite{Jee_2011} for HCS.}


We selected from the WINGS spectroscopic dataset all those clusters with total masses $M_{DM} \geq 5\times10^{14} M_\odot$, or velocity dispersions $\sigma > 670$ km/s. This mass cut assured that only the clusters with total masses greater than or equal to the mass predicted in cosmological simulations of structure formation (e.g.\ \citealt{Fakhouri_2010, Chiang_2013}) for the descendant of the least massive HCS cluster (i.e.\ RCS2345) were considered in the analysis. This selection produced a sub-sample of 29 galaxy clusters.

\subsection{The Field Comparison Sample}

We use the COSMOS/UltraVISTA sample \citep{Muzzin_2013} to build a subsample of red sequence, passive galaxies in the field at redshift $0.8<z<1.5$. UltraVISTA \citep{McCracken_2012} is a survey carried out with the VISTA InfraRed CAMera (VIRCAM) on the Visible and Infrared Survey Telescope for Astronomy (VISTA) at the ESO/Paranal Observatory. The survey covers the $\sim 2$ square degree field of the Cosmic Evolution Survey (COSMOS, \citealt{Scoville_2007_COSMOS}), delivering deep images in the NIR Y, J, H, and Ks filters with image quality FWHM in the range $0.75'' - 0.82''$ and $5 \sigma$ magnitude depth in the range 23-24 mag in all bands. The COSMOS/UltraVISTA catalogue collects all the photometric and spectroscopic data available in the COSMOS field at UV, optical, and IR wavelengths. This sample, which is discussed in \cite{Muzzin_2013}, provides multiwavelength photometry (far UV to far IR) for 216,268 sources. Accurate photometric redshifts ($z_{phot}$), stellar masses, and rest-frame $U$, $V$, and $J$ photometry obtained with SED fitting are also available in the dataset. In this work we use the Ks-selected catalogue which can be downloaded from the UltraVISTA web repository{\footnote{\url{http://www.strw.leidenuniv.nl/galaxyevolution/ULTRAVISTA/Ultravista/K-selected.html}}}. This sample is 90\% complete at $Ks = 23.4$ mag. The selection of the red sequence sample in the UltraVISTA field is discussed in Section 4.3.


\section{Photometry and Cluster Membership}

\subsection{Object Detection and PSF Modelling}

Tables \ref{table1} and \ref{table2} summarise the global properties and observations of the HCS sample. We followed the procedures described in \cite{Cerulo_2014} to detect objects in each image and model the PSF. In summary, we used a modified version of the GALAPAGOS code \citep{Haeussler_2007} to run SExtractor \citep{Bertin_1996} in high dynamic range mode \citep{Rix_2004}. This allowed us to detect the faintest objects in the images with a reliable deblending of the sources in the cores of the clusters. GALAPAGOS runs SExtractor twice, the first time using a configuration setting optimised for the detection of bright objects ({\ttfamily{COLD}} run) and the second time adopting a configuration setting optimised for the detection of faint objects ({\ttfamily{HOT}} run). When the two individual runs are completed, the software merges the catalogues rejecting the double detections from the sample. The ACS and WFC3 observations targeted the clusters in more than one band and thus, for these datasets, we could run SExtractor in dual image mode. We performed the detection on the F850LP images for the ACS fields and on the F110W images for the WFC3 fields and used the other images for measurement. Although four of the clusters have HAWK-I data in both the J and Ks bands, we did not run SExtractor in dual image mode for these images because the sizes of the fields are slightly different.

The PSF was modelled with PSFExtractor (PSFEx) Version 3.9 \citep{Bertin_2011} in all the images, and the PSF FWHM are reported for each image in Table \ref{table2}. We built the multiband photometric catalogues for each cluster by matching the PSF of the single images to the broadest PSF. These catalogues were used to estimate the stellar masses of red sequence galaxies with the {{\ttfamily{lephare}} code \citep{Arnouts_1999, Ilbert_2006}. Stellar masses of HCS red sequence galaxies will be discussed together with galaxy morphology in Cerulo et al. 2015b.

Following \cite{Meyers_2012}, galaxy colours were measured on the images obtained by convolving each image by the PSF of the other image in the filter pair used for the study of the colour-magnitude diagram ({\itshape{cross-convolution}}). Thus, for example, in the cluster RX0152 the F775W image was convolved by the PSF of the F625W image, and the F625W image was convolved by the PSF of the F775W image. This allowed us to correct for PSF differences between the images, avoiding the matching with the broadest PSF in the sample and the consequent reduction in image quality and depth.

\begin{figure*}
  \centering
	\includegraphics[width=0.9\textwidth, trim=0.0cm 0.0cm 0.0cm 0.0cm, clip]{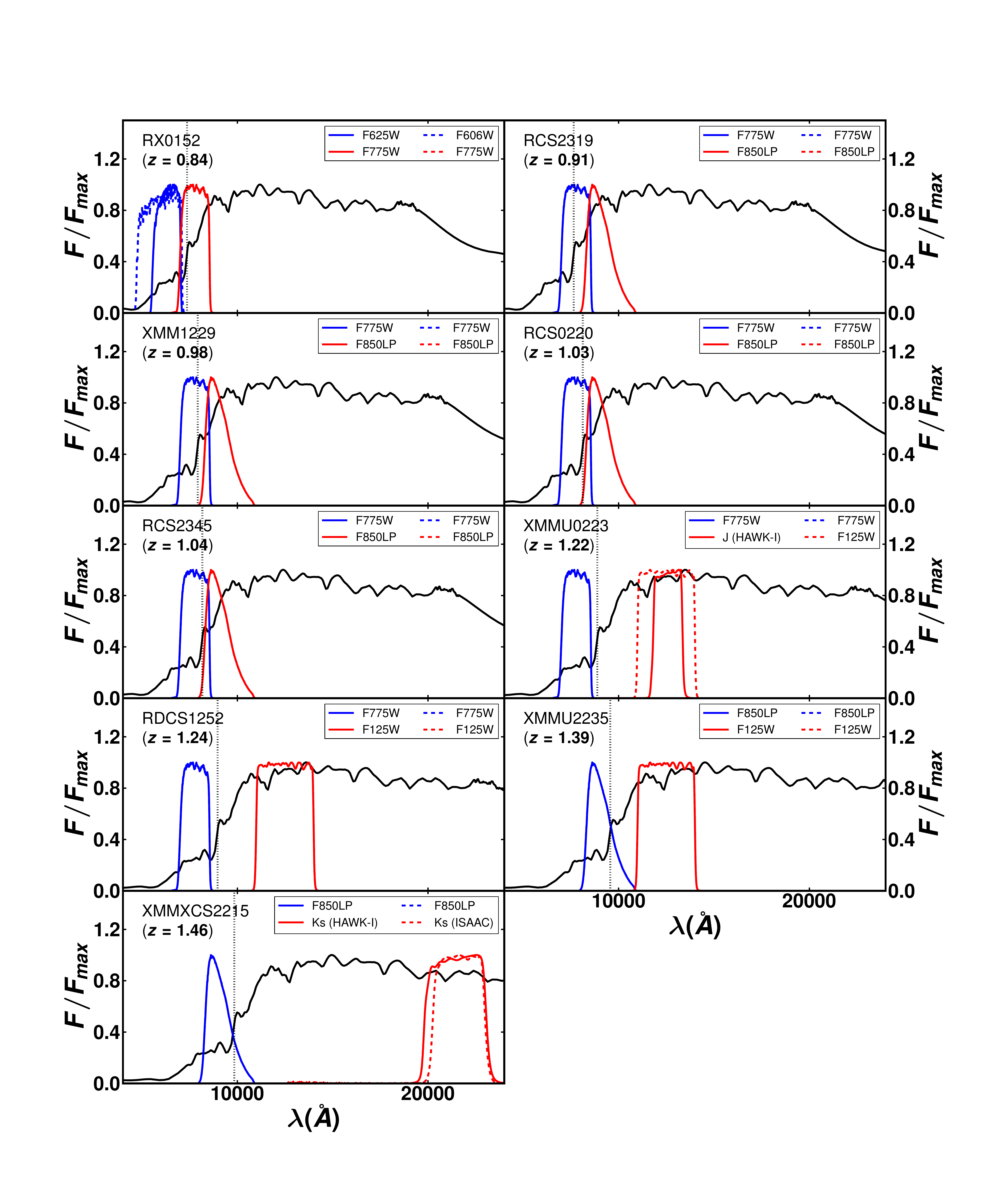}
	\caption{Filter combinations adopted for the study of the red sequence in the HCS clusters. Clusters are ordered by increasing redshift. The solid black line is a template spectral energy distribution of an elliptical galaxy from the library of \protect\cite{Coleman_1980}. The blue and red solid lines are respectively the blue and red filters adopted for the cluster red sequence. The blue and red dashed lines are the blue and red filters adopted in the estimation of field contamination in the GOODS-N/S fields. The names of the filters are written in each plot together with the names and redshifts of the clusters. The vertical dashed lines represent the positions of the 4000 \AA\ break at the redshifts of the clusters. Since no data were available in the ACS F625W band in any of the GOODS fields, we used the F606W GOODS images, which overlap in spectral coverage with the F625W band, in the estimate of background contamination in the RX0152 field (top-left panel). Alternative filters had to be used also for the clusters XMMU0223 (third from the top row, right panel) and XMMXCS2215 (bottom-left panel).}
\label{fig:fig1}
\end{figure*}

\begin{table*}
  \caption{Photometric set-up used for the colour-magnitude diagram and for the estimate of $P_{field}$ in each HCS cluster.}
   \begin{tabular}{|c|c|c|c|}
  \hline
  \multicolumn{4}{|c|}{ } \\
     \multicolumn{1}{|c}{Cluster Name}  & \multicolumn{1}{c}{redshift ($z$)} & \multicolumn{1}{c}{Filter Bands (cluster)}  & \multicolumn{1}{c|}{Filter Bands (field)} \\
     \multicolumn{4}{|c|}{ } \\
      \hline
      \hline
      RX0152      & $0.84$ & $r_{625}$, $i_{775}$     & $V_{606}$, $i_{775}$   \\
      RCS2319     & $0.91$ & $i_{775}$, $z_{850}$     & $i_{775}$, $z_{850}$   \\
      XMM1229     & $0.98$ & $i_{775}$, $z_{850}$     & $i_{775}$, $z_{850}$   \\
      RCS0220     & $1.03$ & $i_{775}$, $z_{850}$     & $i_{775}$, $z_{850}$   \\ 
      RCS2345     & $1.04$ & $i_{775}$, $z_{850}$     & $i_{775}$, $z_{850}$   \\
      XMMU0223    & $1.22$ & $i_{775}$, $J_{HAWK-I}$   & $i_{775}$, F125W      \\
      RDCS1252    & $1.24$ & $i_{775}$, F125W         & $i_{775}$, F125W      \\
      XMMU2235    & $1.39$ & $z_{850}$, F125W         & $z_{850}$, F125W       \\
      XMMXCS2215  & $1.46$ & $z_{850}$, Ks (HAWK-I)   & $z_{850}$, Ks (ISAAC)  \\
  \hline
  \end{tabular} 
\label{table8}
\end{table*}

\subsection{Background Contamination}

In \cite{Cerulo_2014} we estimated cluster membership for XMM1229 using photometric redshifts. However, that sample has the best wavelength coverage in the HCS, with data in 7 passbands from R to Ks, sampling the rest-frame range $0.33 \mbox{ } \mu m < \lambda < 1.11 \mbox{ } \mu m$, and allowing a reliable estimation of photometric redshifts in the cluster field. Most of the HCS clusters have data in only 4 photometric bands and the resulting photometric redshifts are significantly less accurate than those obtained for XMM1229. Thus we decided to estimate the membership of all the HCS clusters, including XMM1229, with the statistical background subtraction technique presented in \cite{Cerulo_2014}. The conclusions of that work regarding the properties of the XMM1229 red sequence remain unchanged.

The filter pairs for the study of the individual red sequences were chosen so that they  bracketed the 4000 \AA\ break at the redshift of the clusters. This allowed us to minimise the contamination from galaxies in the blue cloud. We achieved this goal in all the HCS clusters except RCS2319 ($z=0.91$), XMM1229 ($z=0.98$), RCS0220 ($z=1.03$), and RCS2345 ($z=1.04$), for which we used the ACS F775W and F850LP filters. As shown in Figure \ref{fig:fig1}, to varying degrees, part of the F775W filter lies redward of the 4000 \AA\ break. However, while for the last 3 clusters more than half of the F775W transmission curve covers rest-frame wavelengths $\lambda < 4000$ \AA, in the case of RCS2319 more than 50\% of the transmission curve of the F775W filter falls at wavelengths $\lambda > 4000$ \AA. As a result, the red sequence is flatter than in other clusters at similar redshifts (Figure \ref{fig:fig3a}, Table \ref{table9}). The photometric bands used for the study of the cluster red sequences are summarised in Table \ref{table8} and plotted in Figure \ref{fig:fig1} as solid blue and red lines. \textcolor{black}{The study of the red sequence in the observer frame and at rest frame is discussed in Sections 4.1 and 4.2.}

We built control fields for each cluster using the data taken in the same bands in the Great Observatories Origins Deep Survey (GOODS) North and/or South fields (hereafter GOODS-N and GOODS-S, \citealt{Giavalisco_2004}). This was possible for all the clusters except RX0152, XMMU0223, and XMMXCS2215. For the first cluster we had to resort to the GOODS-N and S ACS F606W ($V_{606}$) images, which overlap with the spectral region sampled by the F625W band (Figure 1), while we used the Cosmic Assembly Near-infrared Deep Extragalactic Legacy Survey (CANDELS, \citealt{Grogin_2011, Koekemoer_2011}) WFC3 F125W images of the GOODS-S field for XMMU0223. \textcolor{black}{We finally used the available ISAAC Ks-band images of the GOODS-S field \citep{Retzlaff_2010} for background subtraction in the XMMXCS2215 field.} The filter pairs used for background estimation in each cluster are shown in Table \ref{table8}, and their transmission curves are plotted in Figure \ref{fig:fig1} as dashed blue and red lines.

The GOODS images were downloaded from the dedicated survey repositories\footnote{CANDELS: {\url{http://candels.ucolick.org/data\_access/Latest\_Release.html}}}\footnote{ISAAC ESO/GOODS: {\url{http://www.eso.org/sci/activities/garching/projects/goods.html}}}, and the source detection and photometry were performed following the same procedure adopted for the cluster fields. 

The PSF of the ISAAC Ks band images varies considerably across the field, with a FWHM spanning the range $0.38 < FWHM < 0.58$. Hence, prior to analyse these images, we matched the PSFs of each image section to that of the image section  with the broadest PSF. 


Following the approach adopted in the analysis of XMM1229, we divided the region of the colour-magnitude diagram where the red sequence is situated in two rectangular cells, respectively corresponding to the regions brighter and fainter than the red sequence magnitude mid point. Then, following \cite{Pimbblet_2002} and \cite{Valentinuzzi_2011}, we computed the field contamination probability, that is, the probability for a galaxy to belong to the field as:
\begin{equation}
P_{field} = \frac{N_{field} \times A - N_{cluster, cont}}{N_{cluster} - N_{cluster, cont}}
\end{equation}
in each colour-magnitude cell, where $N_{field}$ is the number of galaxies in each GOODS-N/S colour-magnitude cell, $N_{cluster}$ and $N_{cluster, cont}$ are respectively the number of galaxies and the number of spectroscopic interlopers in each colour-magnitude cell in the cluster field, and $A$ is the ratio between the areas of the cluster and of the GOODS-N/S fields. 

\textcolor{black}{The right-hand panels of Figure \ref{fig:fig3a} show the plots of the colour-magnitude diagrams in a region randomly chosen in the GOODS fields with surface equal to that considered in the study of the clusters (see Section 4.1).} Also plotted are the best-fit lines and the boundaries of the cluster red sequences with the bright and faint magnitude limits. A qualitative comparison with the cluster colour-magnitude diagrams plotted in the left-hand panels of Figure \ref{fig:fig3a} shows that the field contamination in the cluster fields is globally low and mainly affects the faint end of the red sequence. We find that the field contamination probability spans the range $0.01 < P_{field} < 0.20$ (Table \ref{table9}).

\begin{figure*}
  \centering 
	\subfloat[]{\label{fig:a}}\includegraphics[width=0.7\textwidth, trim=0.0cm 2.5cm 0.0cm 4.0cm, clip, page=1]{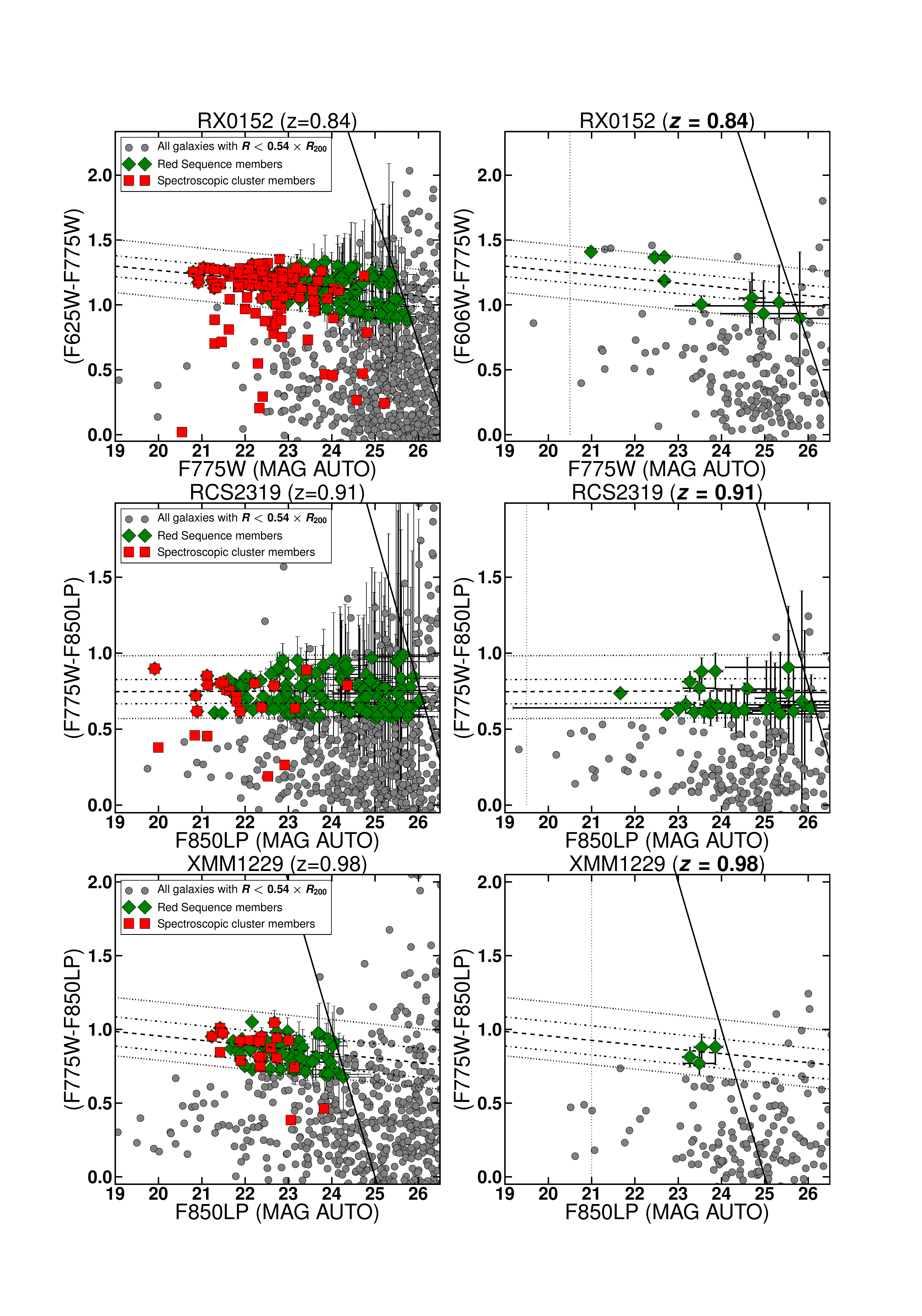}
	\caption{{\itshape{Left}}: Observed colour-magnitude diagrams of the HCS clusters within $0.54 \times R_{200}$ of the cluster centroid. Grey points are all the galaxies observed in the region, green diamonds are galaxies on the observed red sequence and red squares are the spectroscopically confirmed cluster members. The black dashed line is the best-fit straight line to the observed red sequence and the dotted parallel lines represent the red sequence envelope determined as discussed in \S 4.1. The diagonal solid lines correspond to the 90\% completeness limit while the dot-dashed diagonal lines represent the boundaries of the red sequence with the alternative selection $\left|\Delta C  \right| < 3\sigma_{22}$ (see Section 4.1). {\itshape{Right}}: Observed colour-magnitude diagrams in the GOODS-N/S control fields used for field subtraction. Galaxies within a projected spatial region with the same area of that considered for the clusters are plotted in each figure. The fit and boundaries of the observed cluster red sequences and the 90\% magnitude completeness limit are also plotted. The vertical dotted lines represent the apparent magnitude of the brightest red sequence galaxy in each cluster. Galaxies falling within the magnitude and colour ranges of the observed cluster red sequences are plotted as green diamonds. It can be seen that there are few field galaxies with magnitudes and colours in the ranges of the observed cluster red sequences. As a result, the field contamination of the cluster red sequence is low (see also Table \ref{table9}).}
\label{fig:fig3a}
\end{figure*}

\begin{figure*}
  \ContinuedFloat 
  \centering 
  \subfloat[]{\label{fig:b}}\includegraphics[width=0.7\textwidth, trim=0.0cm 2.5cm 0.0cm 4.0cm, clip, page=2]{plot_HCS_and_GOODS_CMD}
  \caption{Continued.}
\label{fig:fig3b}
\end{figure*} 

\begin{figure*}
  \ContinuedFloat 
  \centering 
  \subfloat[]{\label{fig:c}}\includegraphics[width=0.7\textwidth, trim=0.0cm 2.5cm 0.0cm 4.0cm, clip, page=3]{plot_HCS_and_GOODS_CMD}
  \caption{Continued.}
  \label{fig:fig3c}
\end{figure*}

\section{The Red Sequence of the HCS Clusters}

\subsection{The Fitting Procedure}

We studied the red sequence within $0.54 \times R_{200}$ of the projected cluster centre. The choice of this region was imposed by the size of the WFC3 images, which have the smallest field of view in the dataset. However, studying galaxy properties within $\sim 0.5 \times R_{200}$ of the cluster centre allows us to reduce contamination from field galaxies and to easily compare with WINGS, which has a higher spectroscopic sampling towards the cluster centre. In order to quantify the effect of field contamination on the red sequence zero-point, slope and intrinsic scatter, we studied the red sequence in two consecutive steps: (1) we fitted the observed red sequence and obtained first guesses of the parameters with the uncertainties due to photometric error; (2) we fitted the field-corrected red sequence and evaluated the contribution of the interloping objects on the estimates of the fit parameters.

In the first step we fitted a straight line to the observed red sequence by applying a robust line fitting technique based on the Tukey's bi-square weight function \citep{Numerical_Recipes_cpp}. We considered all the galaxies down to the 90\% magnitude completeness limit (diagonal solid lines in Figure \ref{fig:fig3a}) estimated by measuring the fraction of recovered simulated galaxies inserted in random empty regions of the images. The functional form of the colour-magnitude relation is:
\begin{equation}
C_{RS} = a + b \times (m-21.0)
\end{equation}
where $m$ is the apparent magnitude, $a$ is the zero-point, that is, the observer-frame colour at $m=21.0$ mag, and $b$ is the slope. $C_{RS}$ is the galaxy colour on the red sequence in the observer frame. The uncertainties on $a$ and $b$ were estimated as half the width of the 68\% confidence intervals of the distributions of the two parameters after 1000 bootstrap runs. The intrinsic scatter of the red sequence $\sigma_c$ was estimated, as in \cite{Cerulo_2014}, following the approach adopted in \cite{Lidman_2004} and \cite{Mei_2009} and consisting in computing the amount of scatter added to the photometric colour error in order to get reduced $\chi^2 = 1.0$. The uncertainty $\delta \sigma_c$ was estimated as half the width of the 68\% confidence interval of the distribution of $\sigma_c$ after 1000 bootstrap runs.

We defined the red sequence as the locus:
\begin{equation} 
-\kappa_{l} \sigma_c < \Delta C < + \kappa_{h} \sigma_c, 
\end{equation}
where $\Delta C = (C-C_{RS})$ is the difference between the observed ($C$) and best-fit ($C_{RS}$) galaxy colours, $\sigma_c$ is the intrinsic scatter of the red sequence, and $\kappa_{l}$ and $\kappa_{h}$ are factors that were estimated by visually inspecting the colour-magnitude diagram as the most suitable to bracket the red sequence.

The choice of the values of $\kappa_{l}$ and $\kappa_{h}$, which are shown in Table \ref{table9}, was not based on quantitative considerations on the shape of the colour distribution along the red sequence and, in order to test the effect of this selection against a selection based on the photometric scatter, as done in \cite{Delaye_2014}, we also applied an alternative selection consisting in assigning to the observed red sequence all the galaxies within $3\sigma_{22}$ of the best-fit straight line. $\sigma_{22}$ is the colour uncertainty on the red sequence at $m=22.0$ mag, which is (except in XMMXCS2215) the typical magnitude of a bright red sequence member. The selection based on the intrinsic scatter is represented by the dotted diagonal lines in Figure \ref{fig:fig3a}, while the selection based on $\sigma_{22}$ is plotted as dot-dashed diagonal lines in the same figure. The effects of this alternative selection on our results will be discussed in Section 5.2.

The observed colour-magnitude diagrams within $0.54 \times R_{200}$ of the centres of the HCS clusters are plotted in the left-hand panels of Figure \ref{fig:fig3a}, where the objects on the observed red sequences are highlighted as green diamonds and the spectroscopically confirmed cluster members are represented as red squares. Clusters are ordered by increasing redshift.

We note that the photometric errors in some clusters become large at $m>23.0$ mag, and when this effect is particularly severe (e.g.\ RCS2319, middle-left panel in Figure \ref{fig:fig3a}a), we exclude those galaxies from the fit to the red sequence. We stress that including such faint galaxies in the modelling of the red sequence does not affect the estimates of the fit parameters.

In the second step we followed the method outlined by \cite{Valentinuzzi_2011} to statistically estimate the cluster membership of red sequence galaxies. We ran 200 Monte Carlo simulations in which, at each iteration, a random number $0 < P_{sim} < 1$ was assigned to each galaxy. This number was compared with the field contamination probability $P_{field}$ and all the galaxies with $P_{sim} > P_{field}$ were retained as cluster members. We fitted the relation in Equation 3 to the selected {\itshape{cluster members}} using the same procedure adopted in the fit to the observed red sequence and estimated the zero-point, slope, and intrinsic scatter of the {\itshape{contamination-free}} red sequence. The median and half of the 68\% confidence interval of the distributions of the red sequence parameters after 200 iterations were respectively taken as the estimate and uncertainty of each quantity. Figure \ref{fig:fig4} shows these distributions in the case of the cluster RCS0220. The results for all the other clusters are summarised in Table \ref{table9} where, for each cluster, we show the fit parameters of the observed and field-corrected red sequences (top and bottom row in each entry, respectively). Figure \ref{fig:fig4} also shows that slope and zero-point are not independent parameters but are anti-correlated, with steeper slopes corresponding to redder zero-points.

The parameters of the fit to the observed and field-corrected red sequences are all consistent within the uncertainties, and we note that the uncertainty due to field contamination is a factor of $\sim 0.2$ lower than the uncertainty due to photometric errors. In the following analysis we will consider the zero-point, slope, and scatter estimated from the fit to the field-corrected red sequence with the uncertainties defined as the sum in quadrature of the bootstrap and Monte Carlo errors. This will allow us to simultaneously take into account the contributions of photometric errors and field contamination in each cluster sample.

\textcolor{black}{We note that in the XMMXCS2215 field $P_{field} \sim 0.01$. This causes the contribution to the uncertainty on the fit parameters due to field contamination to be negligible with respect to the contribution due to the photometric error. The  Monte Carlo simulations return distributions in which 99\% of the fit parameters are equal to the values obtained from the fit to the observed red sequence. This effect is evident in Table \ref{table9} where the uncertainty due to field contamination is 0 for all the three parameters.}

We compared the slopes and zero-points of the the observed red sequence with those derived form Figure 7 of \cite{Delaye_2014} for the clusters RX0152, RCS2319, XMM1229, RCS0220, and RCS2345, for which the same filters were used in the study of the colour-magnitude diagram, and found that, except for the XMM1229 zero-point, our results are consistent with those of \cite{Delaye_2014} to the $3\sigma$ level.

We built a red sequence sample in WINGS by fitting Equation 3 to the $(B-V)$ vs $V$ colour-magnitude relation. In order to keep the same physical spatial extent of HCS, we only considered the spectroscopically confirmed members within $0.54 \times R_{200}$ of each cluster centre. Before performing the fit the observed $B$ and $V$ Vega magnitudes were converted to absolute rest-frame AB B- and V-band magnitudes. For this conversion we used the distance moduli provided in the catalogues and the k-corrections from \cite{Poggianti_1997}. The selection of red sequence galaxies was performed by adopting the same criterion based on visual inspection of the colour-magnitude diagram for the choice of the two factors $\kappa_l$ and $\kappa_h$.

\begin{figure*}
  \centering
	\includegraphics[width=\textwidth, trim=0.0cm 9.0cm 0.0cm 0.0cm, clip]{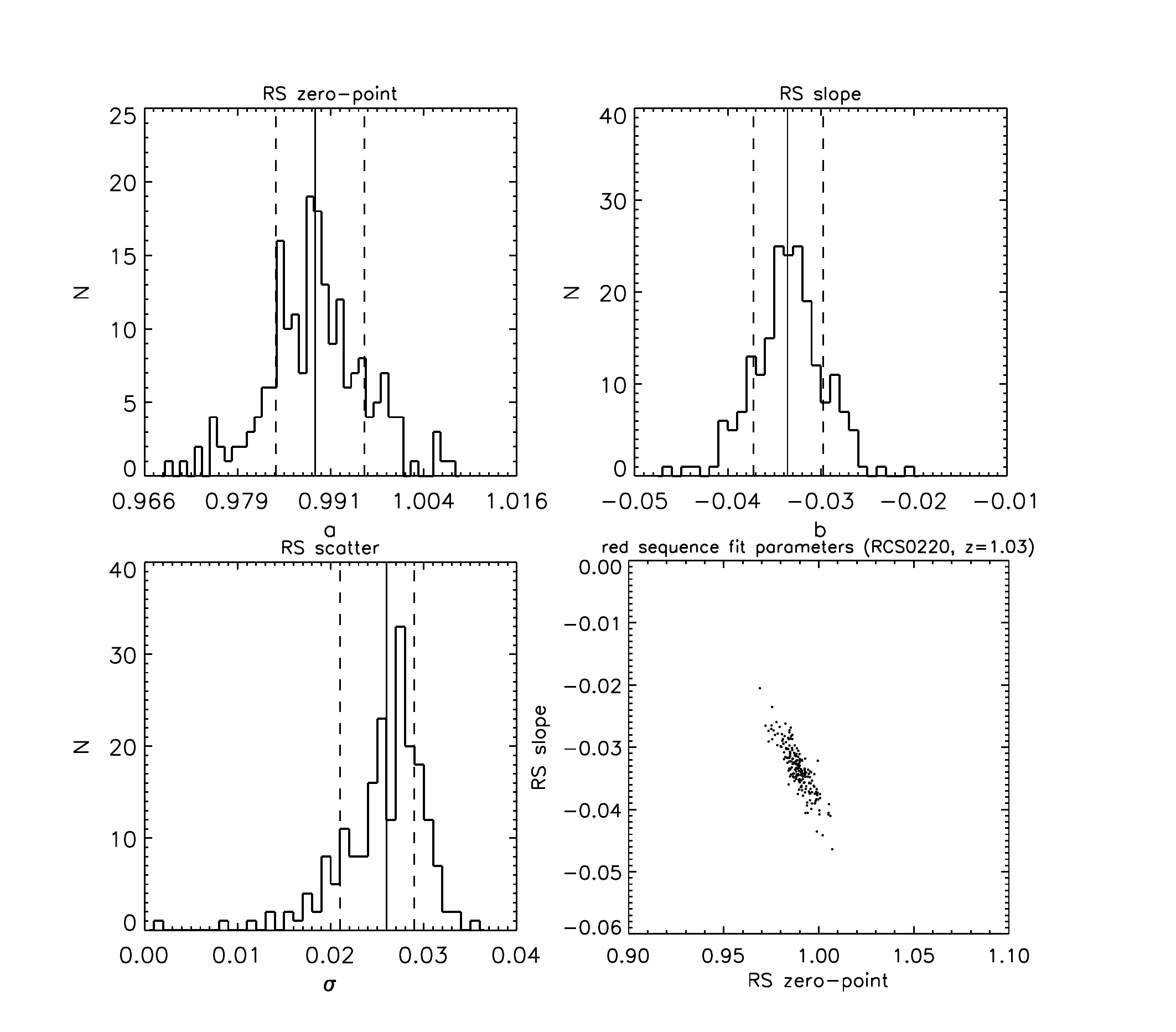}
	\caption{Distribution of the red sequence parameters after 200 Monte Carlo simulations for the statistical estimate of cluster membership. The vertical solid line is the median value of each parameter, while the vertical dashed lines represent the boundaries of the 68\% confidence intervals of the distributions. It can be seen that the red sequence slope and zero-point are highly correlated: steeper slopes correspond to redder zero-points (bottom-right panel).}
\label{fig:fig4}
\end{figure*}

\begin{table*}
  \caption{Parameters of the fits to the observed (top row) and field-corrected (bottom row) red sequences with their respective $1 \sigma$ uncertainties. Also shown are the $\kappa_{l}$ and $\kappa_{h}$ factors adopted in the selection of the red sequence and the values of the probability $P_{field}$ used in the estimation of field contamination. The uncertainties are discussed in \S 4.1.}
  \begin{minipage}{14 cm}
    \begin{tabular}{|c|c|c|c|c|c|c|c|c|}
     \hline
     \multicolumn{7}{|c|}{ } \\
     \multicolumn{1}{|c}{Cluster Name}  & \multicolumn{1}{c}{redshift ($z$)} & \multicolumn{1}{c}{zero-point ($a \pm \delta a$)}  & \multicolumn{1}{c}{slope ($b \pm \delta b$)} & \multicolumn{1}{c}{$\sigma_c \pm \delta \sigma_c$}  & \multicolumn{1}{c}{$\kappa_{min, max}$} & \multicolumn{1}{c|}{$P_{field}$} \\
     \multicolumn{7}{|c|}{ } \\
     \hline
      \hline
     RX0152      & $0.84$ & $1.23 \pm 0.02$   & $-0.033 \pm 0.012$ & $0.034 \pm 0.009$  & 6,6  & 0.09,0.22 \\
                 &        & $1.234 \pm 0.007$ & $-0.034 \pm 0.004$ & $0.018 \pm 0.0015$ &      &            \\
     \hline
     RCS2319     & $0.91$ & $0.75 \pm 0.02$   & $0.001 \pm 0.013$  & $0.059 \pm 0.007$  & 3,4  & 0.09,0.22  \\
                 &        & $0.765 \pm 0.006$ & $-0.012 \pm 0.003$ & $0.042 \pm 0.002$  &      &             \\
     \hline
     XMM1229     & $0.98$ & $0.93 \pm 0.03$   & $-0.030 \pm 0.015$ & $0.033 \pm 0.012$  & 5,7  & 0.02,0.12   \\
                 &        & $0.942 \pm 0.005$ & $-0.052 \pm 0.005$ & $0.041 \pm0.001$   &      &             \\
     \hline
     RCS0220     & $1.03$ & $0.98 \pm 0.03$   & $-0.031 \pm 0.016$ & $0.032 \pm 0.009$  & 6,9  & 0.11,0.12   \\ 
                 &        & $0.988 \pm 0.006$ & $-0.033 \pm 0.003$ & $0.026 \pm 0.005$  &      &              \\ 
     \hline
     RCS2345     & $1.04$ & $0.94 \pm 0.03$   & $-0.010 \pm 0.019$ & $0.02 \pm 0.02$    & 10,8 & 0.07,0.16    \\
                 &        & $0.967 \pm 0.003$ & $-0.048 \pm 0.003$ & $0.025 \pm 0.003$  &      &              \\
     \hline
     XMMU0223    & $1.22$ & $2.11 \pm 0.05$   & $-0.01 \pm 0.04$   & $0.12 \pm 0.02$    & 3,3  & 0.09,0.09    \\
                 &        & $2.113 \pm 0.004$ & $-0.028 \pm 0.009$ & $0.12 \pm 0.004$   &      &              \\
     \hline
     RDCS1252    & $1.24$ & $2.10 \pm 0.04$   & $-0.05 \pm 0.03$   & $0.116 \pm 0.016$  & 3,3  & 0.03,0.12     \\
                 &        & $2.114 \pm 0.006$ & $-0.076 \pm 0.004$ & $0.097 \pm 0.003$  &      &               \\
     \hline
     XMMU2235    & $1.39$ & $1.46 \pm 0.06$   & $-0.01 \pm 0.03$   & $0.086 \pm 0.019$  & 4,4  & 0.10,0.17     \\
                 &        & $1.47 \pm 0.01$   & $-0.036 \pm 0.009$ & $0.109 \pm 0.009$  &      &               \\
     \hline
     XMMXCS2215  & $1.46$ & $2.67 \pm 0.04$   & $-0.10 \pm 0.15$   & $0.24 \pm 0.04$    & 2,4  & 0.012,0.009   \\
                 &        & $2.67 \pm 0.00$   & $-0.10 \pm 0.00$   & $0.24 \pm 0.00$    &      &                \\
  \hline
  \end{tabular}
\end{minipage}
\label{table9}
\end{table*}

\subsection{The Rest-frame Red Sequence}

We followed the approach discussed in Appendix B of \cite{Mei_2009} to convert from the observer-frame photometry adopted for each HCS cluster to rest-frame Vega U, B, and V bands.

We used a set of synthetic stellar population models from the \cite{Bruzual_2003} spectral library spanning the range in formation redshift $2.0 < z_f < 5.0$, with three metallicity values (i.e.: $0.4 Z_\odot$, $Z_\odot$, $2.5 Z_\odot$), two laws of star formation rate (instantaneous burst and exponentially decaying with e-folding time $\tau = 1$ Gyr), and \cite{Salpeter_1955} IMF. For each of the 43 generated models we extracted the observed and rest-frame colours at the redshift of each cluster and fitted the linear relation 
\begin{equation}
C_{rf} = A + B \times C_{obs}
\end{equation}
for the conversion from observed to rest-frame system. $C_{rf}$ is the rest-frame colour, which can be $(B-V)$, $(U-V)$, or $(U-B)$, while $C_{obs}$ is the colour in the observer-frame system adopted for the study of the red sequence in each cluster (Table \ref{table8}). We also extracted the observed and rest frame magnitudes at the redshift of each cluster for each model and used the equation:
\begin{equation}
M_{rf} = m_{obs} + \alpha + \beta \times C_{obs}
\end{equation}
for the conversion to rest-frame magnitudes. $M_{rf}$ is the rest-frame absolute magnitude at the redshift of the cluster, which can be either $B$ or $V$, and $m_{obs}$ and $C_{obs}$ are respectively the apparent magnitude and colour used in the study of each individual red sequence and shown in Table \ref{table8}.

From Equation 5, it follows that the slope and intrinsic scatter of the rest-frame red sequence can be approximated by the relations:
\begin{eqnarray}
b_{rf} & = & B \times b \\
\sigma_{c, rf} & = & B \times \sigma_{c}
\end{eqnarray}
where the subscript $rf$ stands for the rest-frame equivalent of the slope and intrinsic scatter $b$ and $\sigma_{c}$.

We estimated the colours at $V_{Vega} = -20.5$ mag and $B_{Vega} = -21.4$ mag to compare the evolution of the red sequence zero-point $a_{rf}$ with literature results. The reason for this particular combination of choices is that these are the magnitudes at which the zero-points were measured in the works of \cite{Romeo_2008} and \cite{Mei_2009} which are shown in Figure \ref{fig:fig13}. \cite{Valentinuzzi_2011} estimated the red sequence zero-point in the rest-frame $(B-V)$ vs $V$ red sequences of the WINGS clusters at $V_{Vega} = 0.0$ mag. To make the comparison easier and consider the evolution of the zero-point at a magnitude more typical of cluster red sequence galaxies, we evaluated the $(B-V)$ red sequence colour of the WINGS clusters at $V_{Vega} = -20.5$ mag{\footnote{The parameters of the fit to the WINGS red sequences were kindly provided by B.M. Poggianti (private communication).}}.

\begin{figure*}
	\includegraphics[width=\textwidth]{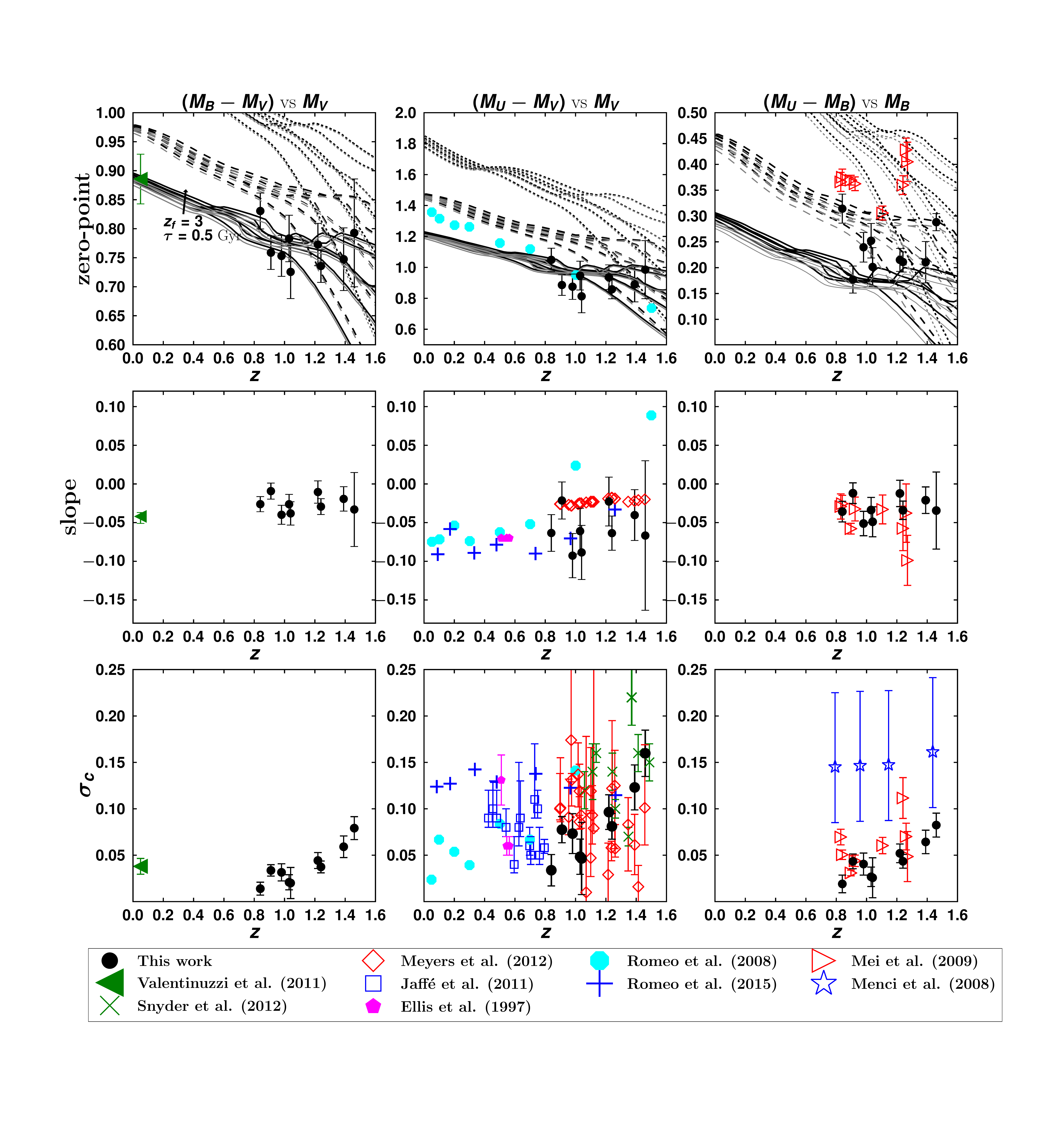}
	\caption{Evolution of the parameters of the rest-frame red sequence and comparison with the literature. From left to right: $(B-V)$ vs $V$ (left column), $(U-V)$ vs $V$ (middle column), $(U-B)$ vs $B$ (right column). From top to bottom: $(B-V)$ and $(U-V)$ colours at $M_V = -20.5$ mag, and $(U-B)$ colour at $M_B=-21.4$ mag (top row), rest-frame slope (middle row), rest-frame intrinsic scatter (bottom row). The hydrodynamical simulations of \protect\cite{Romeo_2008} (cyan octagons) predict a strong evolution of the slope of the red sequence, which becomes positive at $z>1$. This is not in agreement with the observational results of the present work and other works on clusters at $z>0.8$ ( e.g.\ \protect\citealt{Meyers_2012} and \protect\citealt{Mei_2009}). \textcolor{black}{The intrinsic scatter in $(U-V)$ colour exhibits a wide range at $z>0.8$ ($0.01 < (U-V) < 0.25$).} The semi-analytical models of \protect\cite{Menci_2008} predict a large intrinsic scatter, although still consistent with the results of the present work and of \protect\cite{Mei_2009}. The $(U-B)$ colours at $B=-21.4$ mag in \protect\cite{Mei_2009} are systematically redder ($\sim 0.1$ mag) than those estimated in the HCS. As discussed in Section 5.1, those colours were obtained by assuming a common reference redshift $z=0.02$ for all the clusters. All magnitudes are in the Vega system. The lines plotted in the top panels are the colour evolutions derived for a set of models taken from the \protect\cite{Bruzual_2003} library with three different metallicities ($0.4 Z_\odot$-solid, $Z_\odot$-dashed, $2.5 Z_\odot$-dotted), formation redshifts $z_f=3,4,5$, two types of star formation histories (single burst and exponentially declining with $\tau = 0.5$ Gyr) and two IMFs, \protect\cite{Salpeter_1955} (black) and \protect\cite{Chabrier_2003} (grey). \textcolor{black}{The arrow in the top-left plot represents the direction from the model with the lowest formation redshift and longest $\tau$ ($z_f=3.0$, $\tau = 0.5$ Gyr, written in the plot) to that with the highest formation redshift and shortest $\tau$ ($z_f=5.0$, $\tau=0.0$ Gyr). The arrow is only shown in the $(B-V)$ plot and for the models with subsolar metallicity, but the result does not change for the other colours and metallicities.}}
        \label{fig:fig13}
\end{figure*}

Figure \ref{fig:fig13} shows the redshift evolution of the zero-point, slope, and scatter of the $(B-V)_{Vega}$ vs $V_{Vega}$ (left column), $(U-V)_{Vega}$ vs $V_{Vega}$ (middle column), and $(U-B)_{Vega}$ vs $B_{Vega}$ (right column) red sequences. The results of the present work are plotted as black filled circles with error bars corresponding to the $1\sigma$ total uncertainty on the observed red sequence parameters propagated according to the equations:
\begin{multline}
\delta a_{rf} = B \\
                  \times \sqrt{\left[\left(\delta a \right)^2 + \left| \left(m_{ZP} -21.0 \right)\right|^2 \left(\delta b \right)^2 \right]} \\
\end{multline}
\begin{eqnarray}
\delta b_{rf} & = & B \times  \delta b \\
\delta \sigma_{c, rf} & = & B \times  \delta \sigma_{c}
\end{eqnarray}
where $\delta a$, $\delta b$, and $\delta \sigma_{c}$ are the uncertainties on the observed red sequence zero-point, slope, and intrinsic scatter, respectively, and $m_{ZP}$ is the apparent magnitude in the observer frame corresponding to $V_{Vega} = -20.5$ mag or $B_{Vega} = -21.4$ mag. The parameters of the rest-frame red sequences are summarised in Table \ref{table10} and their evolution will be discussed in \S 5.

\begin{table*}
  \caption{Parameters of the fits to the rest-frame red sequence with their respective $1 \sigma$ uncertainties. In each entry: $(B-V)$ vs $V$ (top row), $(U-V)$ vs $V$ (middle row), $(U-B)$ vs $B$ (bottom row). Magnitudes and colours are in the Vega system.}
  \begin{minipage}{10 cm}
    \begin{tabular}{|c|c|c|c|c|}
  \hline
      \multicolumn{5}{|c|}{ } \\
     \multicolumn{1}{|c}{Cluster Name}  & \multicolumn{1}{c}{redshift ($z$)} & \multicolumn{1}{c}{zero-point ($a \pm \delta a$)}  & \multicolumn{1}{c}{slope ($b \pm \delta b$)} & \multicolumn{1}{c|}{$\sigma_c \pm \delta \sigma_c$} \\
     \multicolumn{5}{|c|}{ } \\
     \hline
     \hline
      RX0152     &    $0.84$     &  $0.83 \pm 0.03$   &   $-0.026 \pm 0.010$  &  $0.014 \pm 0.007$ \\
                 &               &  $1.05 \pm 0.08$   &   $-0.06 \pm 0.02$    &  $0.034 \pm0.017$  \\
                 &               &  $0.28 \pm 0.03$   &   $-0.036 \pm 0.013$  &  $0.019 \pm 0.010$ \\
      \hline
      RCS2319    &    $0.91$     &  $0.76 \pm 0.03$   &   $-0.009 \pm 0.010$  &  $0.034 \pm 0.006$ \\
                 &               &  $0.89 \pm 0.07$   &   $-0.02 \pm 0.02$    &  $0.078 \pm 0.014$ \\
                 &               &  $0.15 \pm 0.03$   &   $-0.012 \pm 0.013$  &  $0.044 \pm 0.008$ \\
     \hline
      XMM1229    &    $0.98$     &  $0.75 \pm 0.04$   &   $-0.04 \pm 0.012$   &  $0.032 \pm 0.009$ \\
                 &               &  $0.88 \pm 0.08$   &   $-0.09 \pm 0.03$    &  $0.07 \pm 0.02$   \\
                 &               &  $0.21 \pm 0.03$   &   $-0.051 \pm 0.016$  &  $0.041 \pm 0.012$ \\
      \hline
      RCS0220    &    $1.03$     &  $0.78 \pm 0.04$   &   $-0.026 \pm 0.013$  &  $0.021 \pm 0.008$ \\
                 &               &  $0.95 \pm 0.09$   &   $-0.06  \pm 0.03$   &  $0.048 \pm 0.019$ \\  
                 &               &  $0.25 \pm 0.03$   &   $-0.034 \pm 0.016$  &  $0.027 \pm 0.010$ \\

      \hline
      RCS2345    &    $1.04$     &  $0.73 \pm 0.05$   &   $-0.038 \pm 0.015$  &  $0.020 \pm 0.017$ \\
                 &               &  $0.81 \pm 0.11$   &   $-0.09 \pm 0.03$    &  $0.05 \pm 0.04$   \\
                 &               &  $0.17 \pm 0.04$   &   $-0.049 \pm 0.019$  &  $0.026 \pm 0.02$  \\
      \hline
      XMMU0223   &    $1.22$     &  $0.77 \pm 0.04$   &   $-0.010 \pm 0.014$  &  $0.044 \pm 0.009$ \\
                 &               &  $0.94 \pm 0.08$   &   $-0.02 \pm 0.03$    &  $0.096 \pm 0.019$ \\
                 &               &  $0.18 \pm 0.02$   &   $-0.012 \pm 0.017$  &  $0.052 \pm 0.010$ \\
      \hline
      RDCS1252   &    $1.24$     &  $0.74 \pm 0.03$   &   $-0.029 \pm 0.010$  &  $0.037 \pm 0.006$ \\
                 &               &  $0.86 \pm 0.06$   &   $-0.06 \pm 0.02$    &  $0.081 \pm 0.014$ \\
                 &               &  $0.179 \pm 0.019$ &   $-0.034 \pm 0.012$  &  $0.043 \pm 0.007$ \\
      \hline
      XMMU2235   &    $1.39$     &  $0.75 \pm 0.05$   &   $-0.019 \pm 0.016$  &  $0.059 \pm 0.012$ \\
                 &               &  $0.89 \pm 0.11$   &   $-0.04 \pm 0.03$    &  $0.12 \pm 0.02$   \\
                 &               &  $0.18 \pm 0.04$   &   $-0.021 \pm 0.017$  &  $0.064 \pm 0.013$ \\
      \hline
      XMMXCS2215 &    $1.46$     &  $0.79 \pm 0.09$   &   $-0.03 \pm 0.05$    &  $0.079 \pm 0.012$ \\
                 &               &  $0.99 \pm 0.19$   &   $-0.07 \pm 0.10$    &  $0.16 \pm 0.02$   \\
                 &               &  $0.255 \pm 0.017$ &   $-0.03 \pm 0.05$    &  $0.083 \pm 0.013$ \\
 \hline
  \end{tabular}
\end{minipage}
\label{table10}
\end{table*}

\subsection{The Field Red Sequence Sample}

To build the field comparison sample for HCS, we selected galaxies in the range $0.8 < z_{phot} < 1.5$ and fitted the red sequence in the rest-frame $(U-V)$ vs $V$ colour-magnitude diagram following the same procedure discussed in Section 4.1. As shown in \cite{Muzzin_2013} and \cite{van_der_Burg_2013}, the $U$, $V$, and $J$ photometric bands are highly efficient in discriminating between quiescent and star forming galaxies and, in particular, in correcting for contamination of the red sequence due to dusty star-forming galaxies. By applying the same limits adopted in \cite{van_der_Burg_2013}, who also used the UltraVISTA sample, for quiescent, passively evolving galaxies:
\begin{equation}
(U-V) > 1.3 \cap (V - J) < 1.6 \cap (U - V) > 0.88(V - J) + 0.6
\end{equation}
we found that 35\% of the galaxies selected on the UltraVISTA red sequence are dusty-star-forming galaxies. Thus we used the $UVJ$ selection to clean the field red sequence from these contaminants and adopted the UVJ-selected sample for the study of the luminosity distributions in Section 4.5. Interestingly, after converting the observed magnitudes of the HCS red sequence galaxies to $U$, $V$, and $J$ rest-frame magnitudes, we find that all galaxies selected on the red sequence fall within the quiescent region of the $UVJ$ plane.

\subsection{The Luminous-to-Faint Ratio}

We analyse the build-up of the red sequence as a function of galaxy luminosity by adopting two complementary approaches, namely the study of the evolution of the ratio between luminous and faint galaxies (luminous-to-faint ratio, $L/F$), and the study of the luminosity distribution of galaxies along the red sequence. While the latter will be considered in Section 4.4, in this section we focus on the measurement of $L/F$.

In order to compare the results from different HCS clusters, and to compare between HCS and WINGS, we converted our observer-frame photometry to rest-frame $V_{AB}$ absolute magnitudes as discussed in \S 4.2, using Equations 5 and 6, and passively evolved the rest-frame magnitudes to $z=0$. We considered all red sequence galaxies down to $V_{AB} = -19.5$ mag, which is the 90\% $V_{AB}$ completeness limit of XMMXCS2215, the shallowest sample in our dataset. We defined as luminous all the galaxies with $V_{AB} < -20.5$ mag and as faint all those with $-20.5 \leq V_{AB} < -19.5$, this subdivision corresponding to the mid point of the red sequence magnitude range in XMMXCS2215. The same ranges were adopted in the WINGS sample, and we defined the luminous-to-faint ratio as the ratio $L/F$ between the numbers of luminous and faint galaxies. Figure \ref{fig:fig15} shows that this cut results in the loss of part of the faint galaxy population in deep datasets such as RDCS1252 and RX0152; furthermore, it does not allow a direct comparison with most literature results, which were obtained adopting the ranges $V_{Vega} < -20.0$ mag and $-20.0 < V_{Vega} < -18.2$ for luminous and faint galaxies, respectively (see e.g.: \citealt{De_Lucia_2007, Gilbank_2008, Andreon_2008, Capozzi_2010, Valentinuzzi_2011}). The numbers of luminous and faint galaxies were corrected for field contamination as discussed in Section 3.2.

\textcolor{black}{The estimates of $L/F$ are summarised in Table \ref{table11}, while Figure \ref{fig:fig14} shows the trends of the luminous-to-faint ratio with cluster redshift (left-hand panels) and total mass (right-hand panels). The error bars reported in the plots correspond to the 68\% Poissonian confidence intervals which were estimated adopting the approximations of \cite{Ebeling_2003_Poisson}. The error on the field contamination probability is added in quadrature to the Poisson uncertainty. We studied the contribution of cosmic variance in the two GOODS fields used for the estimate of field contamination, adopting the method presented in \cite{Moster_2011}, and found that the average fractional contribution of cosmic variance in the luminosity/stellar mass range covered by the HCS red sequence amounts to 6\% or 9\% according to whether both fields or only the GOODS-S field was used for background subtraction. Cosmic variance is also added in quadrture to the Poisson and field contamination uncertainties. The error bars on the halo mass of the HCS clusters are taken from Table 2 of \cite{Jee_2011}. The HCS clusters cover the halo mass range $2.0 \times 10^{14} \mbox{ } M_\odot < M_{DM} < 7.5 \times 10^{14} \mbox{ } M_\odot$, going from low-mass, spiral-rich clusters (RCS2345) to very high massive clusters (XMMU0223). The value of the mass reported for WINGS is the median halo mass of the sample, estimated with Equation 1, with the error bars corresponding to the 68\% width of the halo mass distribution. The bottom panels of Figure \ref{fig:fig14} show the plots of $L/F$ as a function of redshift and cluster halo mass for the alternative selection of galaxies on the red sequence with $\left|\Delta C  \right| < 3\sigma_{22}$. These results will be discussed in \S 5.2.}

\begin{table*}
  \caption{Red sequence luminous-to-faint ratio ($L/F$) of the HCS and WINGS clusters. The values of the cluster halo mass from \protect\cite{Jee_2011} are shown in the fourth column from the left. The halo mass and uncertainties quoted for WINGS refer to the median and 68\% confidence interval of the halo mass distribution of the subsample of WINGS clusters used in the comparison with HCS. Halo masses for the WINGS clusters were estimated with Equation 1. The second row in each HCS entry refers to the red sequence selection with $\left| \Delta C \right| < 3\sigma_{22}$ (See Section 4.1).}
  \begin{minipage}{9 cm}
    \begin{tabular}{|c|c|c|c|}
  \hline
      \multicolumn{4}{|c|}{ } \\
     \multicolumn{1}{|c}{Cluster Name}  & \multicolumn{1}{c}{redshift ($z$)} & \multicolumn{1}{c}{$(M_{DM} \pm \delta M_{DM}) \times 10^{14}$ $M_\odot$}  & \multicolumn{1}{c|}{$L/F \pm \delta (L/F)$} \\
     \multicolumn{4}{|c|}{ } \\
     \hline
     \hline
     WINGS       & 0.05   &  $6.0_{-3.0}^{+5.0}$  &  $0.412_{-0.012}^{+0.013}$ \\[7pt]
                 &        &                     &                           \\[7pt]
     RX0152      & 0.84   &  $4.4_{-0.5}^{+0.7}$  &  $0.37_{-0.15}^{+0.20}$    \\[7pt]
                 &        &                     &  $0.6_{-0.3}^{+0.4}$    \\[7pt]
     RCS2319     & 0.91   &  $5.8_{-1.6}^{+2.3}$  &  $0.37_{-0.16}^{+0.20}$       \\[7pt]
                 &        &                     &  $0.26_{-0.16}^{+0.30}$       \\[7pt]
     XMM1229     & 0.98   &  $5.3_{-1.2}^{+1.7}$  &  $0.22_{-0.10}^{+0.14}$     \\[7pt]
                 &        &                     &  $0.31_{-0.14}^{+0.20}$        \\[7pt]
     RCS0220     & 1.03   &  $4.8_{-1.3}^{+1.8}$  &  $0.29_{-0.13}^{+0.19}$        \\[7pt]
                 &        &                     &  $0.5_{-0.2}^{+0.3}$         \\[7pt]
     RCS2345     & 1.04   &  $2.4_{-0.7}^{+1.1}$  &  $1.0_{-0.5}^{+0.6}$          \\[7pt]
                 &        &                     &  $1.5_{-0.8}^{+1.1}$          \\[7pt]
     XMMU0223    & 1.22   &  $7.4_{-1.8}^{+2.5}$  &  $0.6_{-0.3}^{+0.4}$          \\[7pt]
                 &        &                     &  $0.8_{-0.4}^{+0.6}$          \\[7pt]
     RDCS1252    & 1.24   &  $6.8_{-1.0}^{+1.2}$  &  $0.36_{-0.17}^{+0.30}$          \\[7pt]
                 &        &                     &  $0.31_{-0.18}^{+0.30}$        \\[7pt]
     XMMU2235    & 1.39   &  $7.3_{-1.4}^{+1.7}$  &  $0.8_{-0.4}^{+0.6}$          \\[7pt]
                 &        &                     &  $1.0_{-0.6}^{+0.8}$          \\[7pt]
     XMMXCS2215  & 1.46   &  $4.3_{-1.7}^{+3.0}$  &  $0.6_{-0.2}^{+0.3}$          \\[7pt]
                 &        &                     &  $0.7_{-0.3}^{+0.3}$           \\
     \hline
  \end{tabular}
\end{minipage}
\label{table11}
\end{table*}

\subsection{The Red Sequence Luminosity Distribution}

We divided the red sequence of each cluster into 0.5 mag bins in the photometric band used for the colour-magnitude diagram and used the GOODS control fields to estimate $P_{field}$ in each bin as defined in Equation 5. We converted to passively evolved $z=0$ absolute $V_{AB}$ magnitudes and estimated the red sequence number counts in each cluster as:
\begin{equation}
N_{bin} = \sum_{i=1}^{m}{N_{gal, i} \times (1 - P_{field, i})}
\end{equation}
where $N_{gal,i}$ is the $i^{th}$ galaxy in the bin and $P_{field, i}$ is the expected fraction of field interlopers associated with the galaxy; $m$ is the total number of galaxies in each magnitude bin. The number counts in the individual HCS clusters are plotted in Figure \ref{fig:fig15}.

\textcolor{black}{In order to investigate the behaviour of the red sequence number counts with cluster redshift and dark matter halo mass, we created composite samples of clusters at low ($0.8 < z < 1.1$) and high ($1.1 \leq z < 1.5$) redshift and low ($M_{DM} < 5 \times 10^{14} \mbox{ } M_\odot$) and high ($M_{DM} \geq 5 \times 10^{14} \mbox{ } M_\odot$) halo mass and measured the red sequence number counts in each sample. 
Following \cite{Garilli_1999} and \cite{De_Filippis_2011}, in each 0.5 mag bin we estimated the cumulative number counts as:
\begin{equation}
N_{bin} = \frac{1}{M_{bin}}\sum_{i=1}^{m}{\left[N_{gal, i} \times (1 - P_{field, i})\right] w_i^{-1}}
\end{equation}
where $M_{bin}$ is the number of clusters with completeness limits fainter than the considered bin, and $w_i$ is the weight of each cluster defined as the ratio between the number of galaxies in the cluster and the average number of galaxies brighter than its magnitude completeness limit in all the clusters with fainter completeness limits. This method allows one to consider all the clusters in the HCS sample by weighing them according to their completeness limits and enables the study of the luminosity distributions down to the faint end of the red sequence. We adopted the same method to build the red sequence luminosity distribution in the WINGS spectroscopic sample after converting the observed $V$ Vega magnitudes to rest-frame absolute $V$ AB magnitudes and considering only galaxies with $V_{Vega} < 18.0$ mag. As discussed in \cite{Cava_2009}, this magnitude corresponds to 50\% spectroscopic completeness in the WINGS spectroscopic sample. The uncertainties on the number counts were estimated as the width of the 68\% Poisson confidence interval derived as in \cite{Ebeling_2003_Poisson}. The uncertainty on field contamination in HCS and the cosmic variance in the GOODS fields were added in quadrature to the Poisson uncertainties. The HCS and WINGS number counts are plotted as black, filled circles and red diamonds, respectively, in Figures \ref{fig:fig15}, \ref{fig:fig16}, and \ref{fig:fig17}. The WINGS number counts were corrected for spectroscopic incompleteness at each magnitude.}

We stress that the method of \cite{Garilli_1999} is not the only way of combining luminosity distributions. For example, \cite{Colless_1989} used a similar approach but without accounting for the fact that clusters may have different completeness limits. While we will discuss the effect of applying this method to the HCS and WINGS samples in Section 5.2, we refer the reader to \cite{De_Filippis_2011} for a thourough discussion on the application of different methods to a large sample of clusters.

\textcolor{black}{In Figure \ref{fig:fig17} we show the red sequence number counts of the entire HCS sample estimated with Equation 14 plotted against the WINGS red sequence number counts and the UltraVISTA number counts of passive red sequence galaxies at $0.8 < z_{phot} < 1.5$. The error bars in the UltraVISTA number counts include the contribution of cosmic variance estimated using the \cite{Moster_2011} method as done for the HCS number counts and luminous-to-faint ratio. Following the method outlined in \cite{Schmidt_1968}, the number counts in the field were weighted by the maximum comoving volume that each galaxy can occupy according to its luminosity and the magnitude completeness limit of the UltraVISTA sample (i.e. $Ks = 23.4$ mag).}

In Figures \ref{fig:fig15}, \ref{fig:fig16}, and \ref{fig:fig17} are also plotted the best-fit \cite{Schechter_1976} curves expressed by the equation:
\begin{equation}
\Phi(M) = \frac{2}{5} \phi^* \ln{10} \left[10^{0.4(M^*-M)} \right]^{(\alpha+1)} \times e^{10^{0.4(M^*-M)}}
\end{equation}
where $M$ is the absolute magnitude in a given band, $M^*$ is the magnitude at the turn-over point of the luminosity distribution, $\alpha$ is the slope of the faint end of the distribution, and $\phi^*$ is the normalisation. This function is shown to be a good model for the luminosity distribution of galaxies in clusters when brightest cluster galaxies (BCG) are excluded (\citealt{Schechter_1976}, \citealt{De_Filippis_2011}). We fitted Schechter functions to the red sequence luminosity distributions in HCS, WINGS, and UltraVISTA adopting the Maximum Likelihood (ML) technique and assuming that the field-corrected number counts in the two cluster samples follow in each bin a scaled Poisson distribution with expected value equal to the value of the Schechter function at the centre of the bin (see \citealt{Bohm_2014} for a discussion on the use of scaled Poisson distributions for modelling weighted histograms). We excluded from the fit all galaxies at magnitudes brighter than $M_V = -22.4$ mag and $M_V = -23.2$ mag in the HCS and WINGS samples, respectively, as these are the ranges in which the most massive BCGs fall in the two samples. We also excluded the objects in the faintest two bins of the WINGS sample because these are likely to be affected by residual spectroscopic incompleteness. As shown in \cite{Cava_2009}, at $V=18.0$ mag the completeness of the WINGS-SPE sample begins to drop below 50\%.

The fit yields the values shown in Table \ref{table12}, where we report the V-band AB turn-over magnitude $M_V^*$ and the faint-end slope $\alpha$. \textcolor{black}{For each sample we also estimated the Goodness of Fit (GoF) defined as the logarithmic Likelihood Ratio $-2\log\left( \mathcal{L}_{max} / \mathcal{L}_{sat} \right)$, where $\mathcal{L}_{max}$ is the value of the Likelihood function $\mathcal{L}$ corresponding to the parameters of the fit, and $\mathcal{L}_{sat}$ is the value of $\mathcal{L}$ for the saturated model, i.e.\ the model obtained in the limiting case in which the number counts predicted by the fit are equal to the observed number counts}. 

\cite{Andreon_2005_Bayes} show that modelling a distribution after binning the data results in a loss of information. However, for the purposes of this work, in which we are just interested in studying the differences in the trends of the lminosity distributions at the faint end of the red sequence among the three samples and not in an accurate estimate of the Schechter parameters, the modelling of the binned distributions is sufficient.

In all the figures the WINGS and UltraVISTA number counts and Schechter curves are normalised to match the value of the HCS number counts at $M_{V,AB} \sim M_{V,HCS}^*$. The red sequence luminosity distribution will be discussed in \S 5.2.

\begin{table}
  \caption{Results of the fit of the \protect\cite{Schechter_1976} Function to the red sequence luminosity distributions of the HCS, WINGS, and UltraVISTA samples. The fits were performed adopting the Maximum Likelihood method. The $1\sigma$ uncertainties on each parameter ($M_V^*$ and $\alpha$) were obtained by marginalising on the other parameter ($\alpha$ or $M_V^*$) and on the normalisation $\phi^*$. The Goodness of Fit (GoF) is estimated as the Likelihood Ratio $-2\log\left( \mathcal{L}_{max} / \mathcal{L}_{sat} \right)$ (see Section 4.5).}
    \begin{tabular}{|c|c|c|c|}
  \hline
     \multicolumn{1}{|c}{Sample Name}  & \multicolumn{1}{c}{($M_V^* \pm \delta M_V^*$)} & \multicolumn{1}{c}{$\alpha \pm \delta \alpha) $}  & \multicolumn{1}{c|}{GoF} \\
     \hline
     \hline
     HCS         & $-21.08_{-0.02}^{+0.08}$    &  $-0.889_{-0.001}^{+0.009}$   &  20.2       \\
                 &                           &                             &             \\
     WINGS       & $-20.88_{-0.02}^{+0.08}$    &  $-0.91_{-0.04}^{+0.04}$      &  3.0        \\
                 &                           &                             &              \\
     UltraVISTA  & $-21.28_{-0.02}^{+0.08}$    &  $0.22_{-0.02}^{+0.01}$       &  708.0       \\
       \hline
  \end{tabular}
\label{table12}
\end{table}

\section{Discussion}

This section discusses the results presented in Section 4 comparing them with other works published in the literature and framing them in the general context of galaxy evolution.

\subsection{The Evolution of the Red Sequence Parameters}

Figure \ref{fig:fig13} shows the redshift evolution of the HCS red sequence zero-point, slope, and scatter (black circles with error bars), together with the results from the recent literature (coloured symbols) and the predictions for the colour evolution of model stellar populations (lines). The latter were obtained by taking a set of models from the \cite{Bruzual_2003} library with three different metallicities ($0.4 Z_\odot$-solid, $Z_\odot$-dashed, $2.5 Z_\odot$-dotted), formation redshifts $z_f=3,4,5$, two types of star formation history (single burst and exponentially decaying with $\tau = 0.5$ Gyr) and two IMFs, \cite{Salpeter_1955} (black) and \cite{Chabrier_2003} (grey). With these prescriptions for $z_f$ and $\tau$, the models should already be quiescent at $z=1.5$, thus predicting realistic colours for red sequence galaxies.

 A comparison of the red sequence parameters with literature results is difficult because the definitions of quantities such as the intrinsic scatter and the fitting techniques adopted in each work \textcolor{black}{are} different. Differences in the photometric techniques (e.g.\ fixed vs variable aperture photometry), the spectral libraries used in the estimate of the absolute magnitudes, and the adopted cosmologies may also contribute to systematic differences between the results of different works. Furthermore, some authors prefer to apply a morphological selection to their red sequence samples, retaining only elliptical and S0 galaxies. With these caveats in mind we now discuss the evolution of the red sequence colour-magnitude relation in clusters.

The top panels of Figure \ref{fig:fig13} show the evolution of the rest-frame zero-points in the $(B-V)$, $(U-V)$ and $(U-B)$ colours, respectively. The $(B-V)$ red sequence colours at $V_{Vega} = -20.5$ mag are all consistent, within the uncertainties, with the mean value calculated on the red sequence of the WINGS clusters from the \cite{Valentinuzzi_2011} measurements (green triangles). The latter estimate is redder than the HCS zero-points, in agreement with the predictions from the colour evolution of the model stellar populations. We note that this value and the $(B-V)$ zero-points of the HCS clusters are in better agreement with low-metallicity models, as expected from the fact that colours are estimated at $M_V < M_V^*$ in both samples (see Table \ref{table12}). This effect is again evident in the $(U-V)$ zero-points in the middle panel, which are also estimated at $V_{Vega} = -20.5$ mag. The latter zero-points are consistent with the values predicted by the simulations of \cite{Romeo_2008} at $z \sim 1$ (cyan octagons). The $(U-B)$ colours at $B_{Vega} = -21.4$ mag in the right-hand panel are, instead, bluer than those measured in \cite{Mei_2009} for clusters at similar redshifts (red triangles). The clusters RX0152 ($z=0.84$) and RDCS1252 ($z=1.24$) were also in the sample analysed by \cite{Mei_2009} and, while the $(U-B)$ colours are still consistent within the uncertainties for RX0152, the results are discrepant in the case of RDCS1252. We note that \cite{Mei_2009} estimated their $(U-B)$ zero-points at a common redshift $z=0.02$ for all the clusters and used only early-type galaxies to fit the colour-magnitude relation. We also conducted an additional test by running our software to convert the observer-frame zero-points reported in \cite{Mei_2009} to the rest-frame $(U-B)$. We found that the new estimates were $\sim 0.05$ mag bluer than those shown in Figure \ref{fig:fig13}, but still redder than the HCS zero-points. This suggests that additional contributions to this difference may arise from different software implementations of the conversion to the rest-frame, different selection techniques for the red sequence samples, and different photometric strategies (we measure colours within fixed apertures, while \citealt{Mei_2009} used variable apertures). We do not exclude that redder zero-points may be characteristic of some clusters.

The middle panels in Figure \ref{fig:fig13} show the evolution of the rest-frame red sequence slope in all the filter pairs together with the results of other studies and the predictions of the hydrodynamical simulations of \cite{Romeo_2008} and \cite{Romeo_2015}{\footnote{Values of slope and scatter for these simulations were kindly provided by A. Romeo (private communication).}}. It can be seen that our results are consistent with the results of \cite{Meyers_2012} (red squares) and \cite{Mei_2009} in the same redshift range. The sample of \cite{Meyers_2012} includes all the HCS clusters except RX0152 and, interestingly, these authors find shallower slopes than in the present analysis. We attribute this difference to the fact that \cite{Meyers_2012} measured galaxy colours within variable apertures equal to 1 half-light radius, which, as shown by \cite{Scodeggio_2001}, results in systematically shallow slopes. The HCS $(U-B)$ vs $B$ red sequence slopes are consistent with those measured by \cite{Mei_2009}. \cite{Romeo_2008} predicted positive slopes at $z>1$ (middle panel), which does clearly not agree with any of the observational results shown in the plots. However, the predictions of the upgraded hydrodynamical simulations of \cite{Romeo_2015} (blue {\itshape{plus}} symbols) are in good agreement with observational results. We note, indeed,  that the observed rest-frame slopes at $z>1$ are consistent with those measured at lower redshift by \cite{Ellis_1997} (magenta pentagons) and \cite{Valentinuzzi_2011}.

The bottom panels in Figure \ref{fig:fig13} show the evolution of the rest-frame scatter. We see that our results are consistent with \cite{Valentinuzzi_2011} (left-hand panel) although clusters at $z>1.1$ in HCS tend to have slightly higher scatters. Our results are also consistent with the observations of \cite{Meyers_2012}, \cite{Snyder_2012} (green crosses), and \cite{Mei_2009} at similar redshifts, and with the results of \cite{Ellis_1997} and \cite{Jaffe_2011} (blue squares) at $0.4 < z < 0.8$. Our measurements also agree with the predictions of the hydrodynamical simulations of \cite{Romeo_2008} and \cite{Romeo_2015}{\footnote{Except in the case of RX0152, which has a significantly lower scatter than that predicted at similar redshifts by \cite{Romeo_2015}.}} (middle panel) and of the semi-analytical models of \cite{Menci_2008} (right-hand panel, blue stars), although all these works predict higher intrinsic scatters. In particular, \cite{Menci_2008} predict the average $(U-B)$ intrinsic scatter $\sigma_{c,UB} \sim 0.14$ mag, which is higher than the values measured in HCS and in \cite{Mei_2009}.

The results of this study and the comparison with previous works support the notion of an early assembly of the cluster red sequence, which at $z=1$ has already a negative slope and an intrinsic scatter consistent with those measured at lower redshift. This is in agreement with the recent cosmological simulations of \cite{Gabor_2012}, \cite{Romeo_2015}, and \cite{Merson_2015_arxiv} but not with the earlier theoretical works of \cite{Menci_2008} and \cite{Romeo_2008}. The latter authors predicted either a flat non-evolving red sequence \citep{Menci_2008} or a strongly evolving red sequence which, at $z>0.8$, is expected to have a positive slope and a large intrinsic scatter (\citealt{Romeo_2008}). In the latter work the brightest galaxies reach the red sequence last because at $z>1$ they are still building up their stellar mass through gas-rich mergers inducing starbursts. Though this is not observed in most clusters at $z>1$, we note that \cite{Fassbender_2014} found that the brightest galaxy in the core of the cluster XDCP J0044.0-2033, at $z=1.58$, shows signs of recent star formation in its spectrum (post-starburst) and is bluer than the red sequence. A constant or slowly evolving slope supports the notion of the red sequence being primarily a metallicity sequence, with the most massive galaxies being also the most metal-rich systems \citep{Kodama_1997}.  

\cite{Gabor_2012} were able to predict a red sequence with negative slope at $z \sim 1$ by implementing a feedback mechanism based on the interplay between the galaxy and its host halo. When the halo mass crosses a critical value ($10^{12.5} M_\odot$), virial shocks are triggered, heating the halo gas, which can no longer fall into the galaxy and fuel star formation. The critical halo mass corresponds to a stellar mass $M_\star \sim 10^{10.5} M_\odot$ at $1.0 < z < 2.0$ \citep{Moster_2010}, which is typical of bright red sequence galaxies in the HCS. The build-up of the red sequence, therefore, begins at the high mass end, and the brightest and most metal-rich galaxies are the first to join the red sequence, producing the negative slope observed in most high redshift clusters.

The simulations of \cite{Romeo_2008}, \cite{Romeo_2015}, and \cite{Menci_2008} also predict high red sequence scatters at $z>0.8$, higher than those observed in clusters at $z>1$. However, while in \cite{Romeo_2008} $\sigma_c$ increases with redshift, in agreement with the evolution of synthetic stellar populations (see upper panels in Figure \ref{fig:fig13}), \cite{Menci_2008} and \cite{Romeo_2015} predict a flat redshift trend. Our results suggest that a mild increase in the intrinsic scatter may exist at $z>1$. However, given the difficulties inherent in the measurement of this quantity, we would need a larger sample of clusters at $z>1$ to test this scenario. Furthermore, the estimates of $\sigma_c$ in the HCS clusters are all consistent within the uncertainties, thus not ruling out a constant trend resulting from a fast build-up of the cluster red sequence. We also stress that the measurement of the red sequence scatter is strictly related to the particular red sequence sample. We selected the red sequence by considerig all galaxies down to the 90\% magnitude limit in each cluster (Figure \ref{fig:fig3a}), implying that we compare values of $\sigma_c$ obtained within a variable flux limit. While this maximises the information that can be recovered from each cluster, it may introduce artificial trends due to the fact that one compares intrinsically different galaxy populations (e.g.\ bright galaxies in a cluster with faint galaxies in another cluster). When we derive the intrinsic scatter within the same absolute magnitude limit for all the clusters, we find that the trend of $\sigma_c$ with redshift is flat.

\begin{figure*}
	\includegraphics[width=0.8\textwidth, trim=0.0cm 0.0cm 0.0cm 0.0cm, clip]{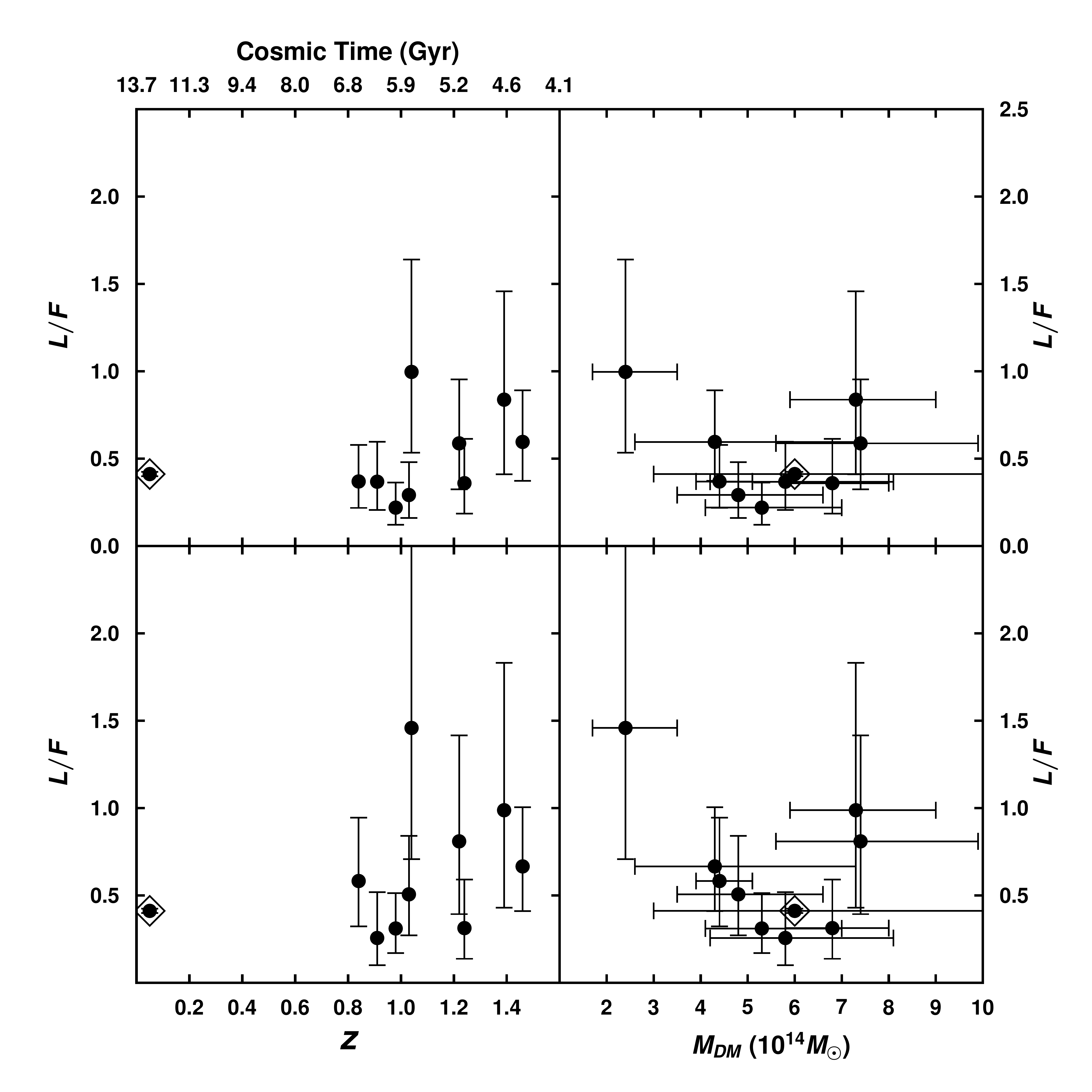}
	\caption{The luminous-to-faint ratio ($L/F$) of the cluster red sequence in the HCS. {\itshape{Top-left panel}}: redshift evolution of $L/F$. {\itshape{Top-right panel}}: $L/F$ as a function of cluster halo mass $M_{DM}$. The diamonds correspond to $L/F$ of the composite WINGS spectroscopic sample. The $M_{DM}$ estimate for WINGS corresponds to the median halo mass estimated with Equation 1, while the error bars correspond to the width of the 68\% confidence interval of the $M_{DM}$ distribution. The bottom panels show $L/F$ as a function of cluster redshift (left) and cluster halo mass (right) for the alternative selection of galaxies on the red sequence with $\left| \Delta C \right| < 3\sigma_{22}$. \textcolor{black}{With both selections, we find little evolution of $L/F$ with redshift and little variation with cluster halo mass.}}
        \label{fig:fig14}
\end{figure*}

\subsection{The Luminous-to-Faint Ratio and the Luminosity Distribution of the Cluster Red Sequence}


Figure \ref{fig:fig14} (top panels) shows the luminous-to-faint ratio $L/F$ plotted as a function of cluster redshift and halo mass. $L/F$ appears constant with both redshift and halo mass, although the distributions are broad in both cases. This result agrees with the conclusions of \cite{Andreon_2008}, \cite{Crawford_2009}, and \cite{De_Propris_2013}, suggesting that no deficit is observed at the faint end of the red sequence in distant clusters. This conclusion does not change if we select galaxies on the red sequence by considering the objects with $\left|C-C_{RS}\right| < 3\sigma_{22}$, where $\sigma_{22}$ is the colour uncertainty on the red sequence at apparent magnitude $m=22.0$ mag (bottom panels). We note that the errors on $L/F$ are large, underlining the high uncertainties related to this quantity and also observed in other works (e.g.: \citealt{De_Lucia_2007}). The flat trend with halo mass (Figure \ref{fig:fig14}, right-hand panels) is in agreement with the conclusions of \cite{De_Lucia_2007}, who found no significant difference between clusters with high ($\sigma > 600$ km/s) and low ($\sigma < 600$ km/s) velocity dispersions at $0.4<z<0.8$ in the ESO Distant Cluster Survey (EDisCS, \citealt{White_2005}).


\begin{figure*}
  \centering
	\includegraphics[width=0.7\textwidth, trim=0.0cm 2.5cm 0.0cm 0.0cm, clip, page=1]{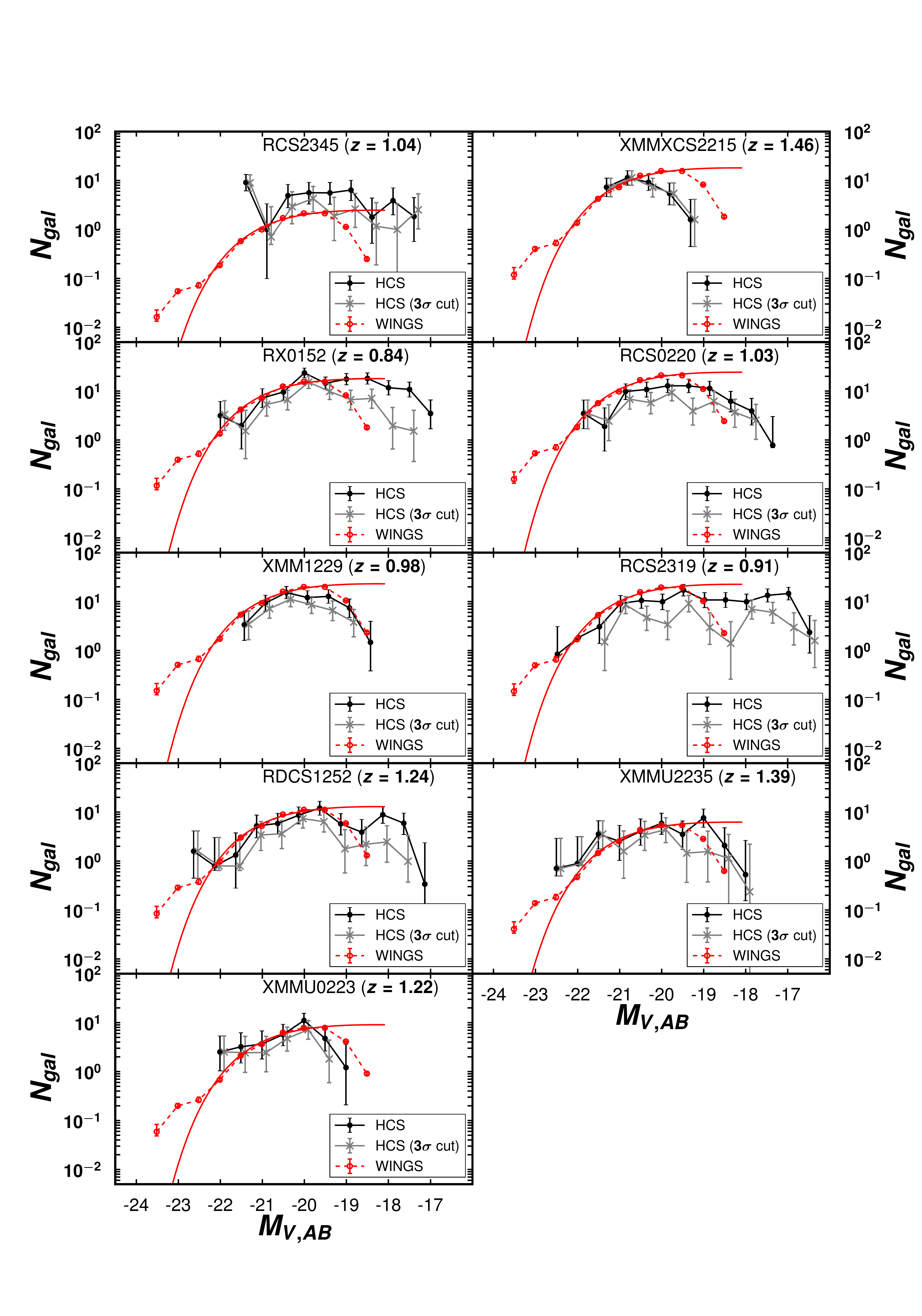}
	\caption{Red sequence number counts in the HCS and WINGS. HCS clusters are ordered by increasing halo mass. Black points and solid connecting lines are for HCS, red circles and dashed connecting lines are for WINGS. The solid red line represents the \protect\cite{Schechter_1976} function fitted to the WINGS red sequence luminosity distribution. Number counts are shown as a function of $V$-band AB absolute magnitude passively evolved to $z=0$. The WINGS number counts and Schechter functions are normalised to match the HCS number counts at approximately $M_V^*$ (see Table \ref{table12}). The grey crosses and the solid connecting grey lines are the HCS red sequence number counts with the alternative selection of red sequence galaxies with $\left| \Delta C \right| < 3\sigma_{22}$. \textcolor{black}{Except for some hints in XMMXCS2215, there is no deficit of galaxies at the faint end of the red sequence. The alternative selection results in a loss of galaxies at faint magnitudes due to the higher photometric error (see Figure \ref{fig:fig3a}), but the number counts remain consistent with those obtained with the initial selection.}}
        \label{fig:fig15}
\end{figure*}

Figure \ref{fig:fig15} shows the luminosity distributions of red sequence galaxies in each individual cluster ordered by increasing halo mass. Except for the last bin, which may start to be affected by magnitude incompleteness, in all clusters the number counts do not deviate significantly from the WINGS composite red sequence. However, XMMXCS2215 ($z=1.46$) shows some hints of a larger deviation at $M_V > M_V^*$, suggesting that \textcolor{black}{the faint end of the red sequence} may still be assembling in this system. We note that a population of star-forming galaxies was discovered in the core of XMMXCS2215 by \cite{Hilton_2010}, supporting the notion that this cluster is in an early stage of its assembly. Interestingly, in the case of XMMU2235 ($z=1.39$), our results agree with \cite{Lidman_2008}, who found no truncation in the red sequence in this cluster. The grey crosses connected by solid lines are the HCS red sequence number counts estimated by adopting the alternative selection $\left|C-C_{RS}\right| < 3\sigma_{22}$. It can be seen that the effect of this selection is a loss of galaxies at faint luminosities, where photometric errors are larger (Figure \ref{fig:fig3a}) and galaxies may be easily scattered off the red sequence. Despite the loss of objects at the faint end, the number counts are still consistent with those obtained with the selection based on the intrinsic scatter.

In the top panels of Figure \ref{fig:fig16} we show the red sequence number counts for the composite samples of clusters at $0.8 < z < 1.1$ (low-z, left-hand panel) and $1.1 < z < 1.5$ (high-z, right-hand panel). We note that there is no significant difference between low- and high-z clusters in the HCS, and no evident truncation of the red sequence with respect to WINGS. This conclusion further supports the scenario of an early build-up of the cluster red sequence. The effect of the alternative selection with $\left|C-C_{RS}\right| < 3\sigma_{22}$ is also in these two cases the decrease in the number counts at faint luminosities, which, however, are still consistent with the number counts obtained from the selection based on the intrinsic scatter.

The central panels in Figure \ref{fig:fig16} show the red sequence number counts in the two low (left-hand panel) and high (right-hand panel) halo mass samples. The low and high mass samples are defined as those containing all clusters with $M_{DM} < 5 \times 10^{14} M_\odot$ and $M_{DM} \geq 5 \times 10^{14} M_\odot$, respectively. This value approximately corresponds to the turn-over point of the halo mass function (see e.g.\ Figures 1 and 2 in \citealt{Jenkins_2001}). The low-mass sample contains the clusters RX0152, RCS0220, RCS2345, and XMMXCS2215, while the high mass sample contains the clusters RCS2319, XMM1229, XMMU0223, RDCS1252, and XMMU2235. 

The bottom panels of Figure \ref{fig:fig16} show the red sequence number counts of the low- and high-mass samples after the removal of RX0152 and RDCS1252. These two clusters present large differences between their halo masses estimated through weak lensing and X-ray luminosity. In particular, as shown in \cite{Delaye_2014}, the X-ray estimated halo masses would cause RX0152 to fall in the high-mass sample, and RDCS1252 to fall in the low-mass sample. We note that, while the red sequence of the massive sample is populated at magnitudes $M_V < -22.0$ mag, no object is detected in the low-mass sample. This result is in agreement with \cite{Lemaux_2012}, who found that the red sequence in the most massive and virialised clusters of the Cl1604 supercluster, at $z=0.9$, is populated at luminosities $\log(L_B/L_\odot) > 10.9$, where less massive systems show a lack of objects. This result suggests that the bright end of the red sequence evolves faster in high-mass clusters. However, we stress that this difference is driven by only 8 galaxies and that the subsamples of low- and high-mass clusters are made of 3 and 4 clusters, respectively, thus not ruling out the hypothesis that the difference between the two luminosity distributions is due to statistical fluctuations.

We also note that the bright end of the WINGS red sequence is populated at $M_V < -22.5$ mag where there is no galaxy detected in the HCS. This range is mainly, but not only, populated by the WINGS BCGs, and our results suggest that these massive galaxies formed at redshifts $z < 0.8$, probably via subsequent dry mergers as suggested in \cite{Faber_2007}. Interestingly, \cite{Lidman_2013} found that BCG growth at $z\sim1$ is driven by major mergers, although accretion of small companions may play an important role in the evolution of these galaxies at lower redshifts \citep{Jimenez_2011}.

\begin{figure*}
  \centering
	\includegraphics[width=0.7\textwidth, trim=0.0cm 2.5cm 0.0cm 0.0cm, clip, page=2]{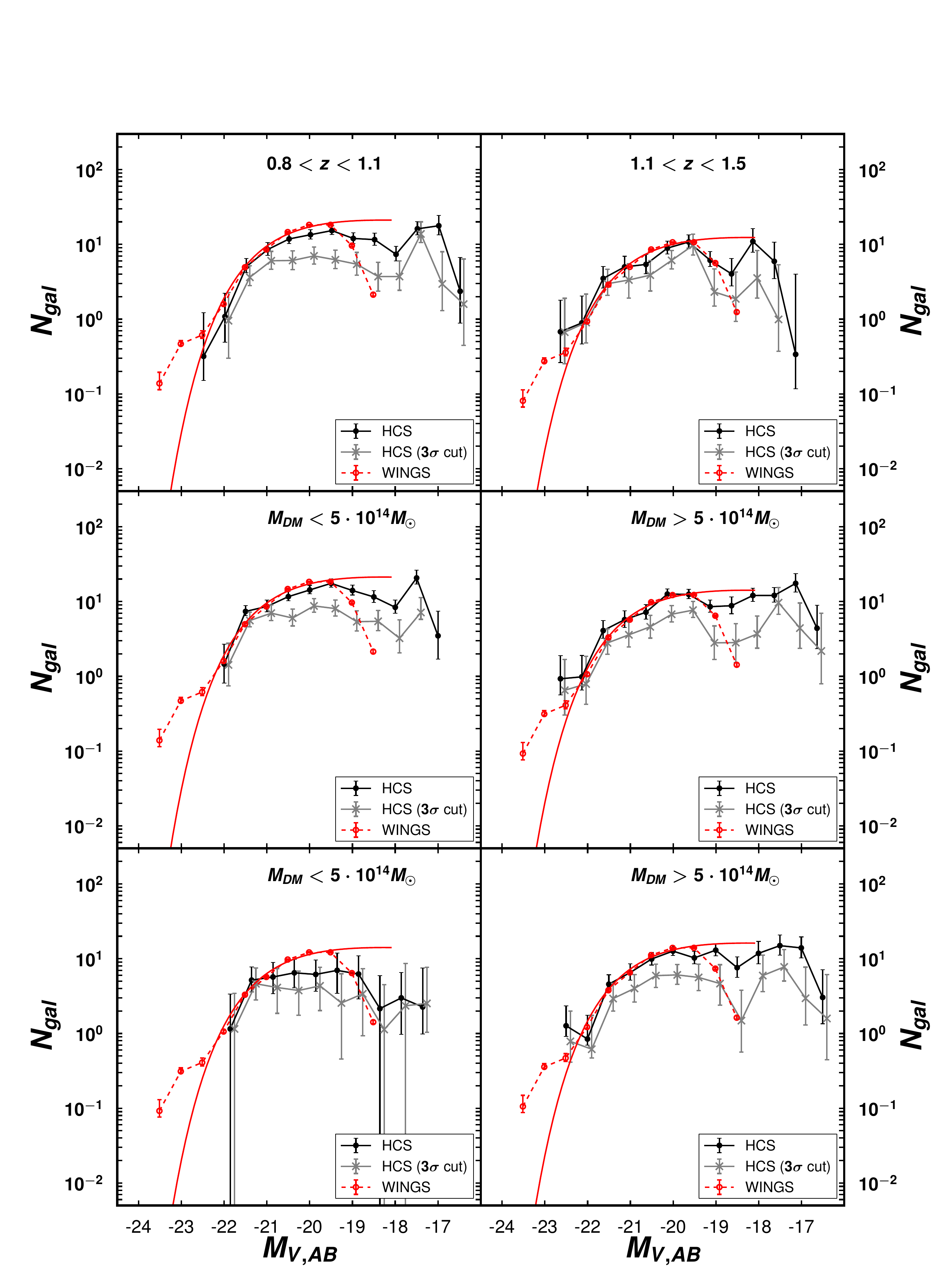}
	\caption{Red sequence number counts at low and high redshifts and low and high cluster halo mass. Colours and symbols are as in Figure \ref{fig:fig15}. {\itshape{Top panels}}: red sequence number counts in clusters at $0.8 < z < 1.1$ (left) and $1.1 < z < 1.5$ (right). {\itshape{Central panels}}: red sequence number counts in clusters with $M_{DM} < 5 \times 10^{14}M_\odot$ (low-mass sample, left) and $M_{DM} \geq 5 \times 10^{14}M_\odot$ (high-mass sample, right). {\itshape{Bottom panels}}: the same as in the middle panels but excluding RX0152 and RDCS1252, the two clusters with large differences between X-ray and weak-lensing halo masses. \textcolor{black}{High halo mass clusters host a population of bright red sequence galaxies with $M_V < -22.0$ mag (bottom-right panel) which are not observed in low-mass clusters. This suggests that the red sequence is more developed at its high mass end in high halo mass clusters.}}
        \label{fig:fig16}
\end{figure*}

The alternative selection of red sequence galaxies with $\left|C-C_{RS}\right| < 3\sigma_{22}$ results also in this case in a loss of galaxies at faint magnitudes. The number counts are, however, still consistent with those obtained from the selection based on the intrinsic scatter.

The faint end of the red sequence does not show any deficit of galaxies in all the subsamples, confirming the result of the luminous-to-faint ratio and suggesting a scenario in which the cluster red sequence is already assembled at $z=1.5$ at the faint end. 

Figure \ref{fig:fig17} shows the red sequence luminosity distribution of the entire HCS sample (black filled circles) plotted against the WINGS red sequence number counts (red diamonds) and the number counts of passive red sequence galaxies in the COSMOS/UltraVISTA field at $0.8 < z_{phot} < 1.5$ (blue crosses). Also plotted are the best-fit Schechter curves for the three samples. Table \ref{table12} shows that, while the HCS and WINGS composite luminosity distributions are well fitted by a Schechter function, the fit performs poorly for the UltraVISTA sample. In particular, we find $GoF=612.1$ with 10 degrees of freedom, which suggests that the Schechter model should be rejected for this dataset. Nonetheless, the aim of this paper is to discuss the differences between the red sequences of different samples at the faint end, and the Schechter $\alpha$ parameter still provides information on the trend followed by galaxy number counts at faint magnitudes. Figure \ref{fig:fig17}b shows the 68\%, 90\%, and 99\% confidence contours of the GoF surfaces, obtained marginalising over $\phi^*$ and using the values derived in \cite{Avni_1976}.

Our estimate of the faint-end slope $\alpha$ in WINGS is consistent with \cite{Moretti_2015} within $2\sigma$, while we find that there is a large difference ($>3\sigma$) between our $M_V^*$ estimate and that obtained by \cite{Moretti_2015}. This difference is exacerbated if we build the composite number counts with the \cite{Colless_1989} method which, unlike the method adopted for this analysis, does not weigth according to the different completeness limits of the clusters. We think that this difference may be due to the fact that we only restrict our analysis to the red sequence, while \cite{Moretti_2015} study the luminosity distributions of both blue and red galaxies. Furthermore, \cite{Moretti_2015} conducted their analysis on the entire WINGS sample and not on a subsample of massive clusters as done in this paper.

It can be seen in Figure \ref{fig:fig17}b that, while the values of $\alpha$ in HCS and WINGS are consistent within the uncertainties, the UltraVISTA sample occupies a statistically different region of the $M^*$-$\alpha$ plane. This suggests that there is no truncation of the red sequence in high redshift clusters, but that there is a deficit at the faint end of the red sequence in the field. 

Before discussing the implications of these results it is important to stress that all methods for building composite luminosity distributions are based on weighting schemes, which may introduce artificial trends. Furthermore, the conclusions on the properties of the composite luminosity distributions obtained with the same method on different samples may be significantly different. We applied the \cite{Colless_1989} method to build the composite luminosity distributions of the HCS and WINGS samples and compared the values of the Schechter parameters with those obtained with the \cite{Garilli_1999} method outlined in Section 4. We found that for HCS the values of $M_V^*$ in the two composite luminosity distributions are consistent, but the estimates of $\alpha$ are discrepant, with the \cite{Colless_1989} method producing a shallower faint-end slope. For WINGS, instead, both parameters are discrepant, and the faint-end slope is significantly steeper. If we compare the $\alpha$ parameters in HCS and WINGS, we find that the HCS faint-end slope is significantly shallower than the WINGS faint-end slope. If one only adopted the \cite{Colless_1989} method in the analysis of these two samples, the conclusion would be that the HCS red sequence is truncated. Nevertheless, both WINGS and HCS would still have a significantly steeper faint-end slope with respect to UltraVISTA.

From Figure \ref{fig:fig17}a it can be seen that the HCS and WINGS data-sets cover two different magnitude ranges, and that only the range $-22.8 < M_V < -18.4$ is covered by all the samples. By running a Kuiper's test (\citealt{Numerical_Recipes_cpp} and references therein) to compare the HCS luminosity distribution with those of WINGS and UltraVISTA we find that there is a probability $P_{Kuiper} < 0.1$\% that the HCS and UltraVISTA luminosity distributions are drawn from the same parent distribution and a probability $P_{Kuiper} = 85$\% that the HCS and WINGS red sequence luminosity distributions are drawn from the same parent distribution. By using the composite luminosity distributions obtained with the \cite{Colless_1989} method we find $P_{Kuiper} < 0.1$\% that HCS and UltraVISTA are drawn from the same parent distribution and $P_{Kuiper} = 51$\% that HCS and WINGS are drawn from the same parent distribution. Although the latter probability is lower than that obtained with the \cite{Garilli_1999} method, it is sufficient to rule out that the two distributions are different. The Kuiper's test is an extension of the Kolmogorov-Smirnov test and, although both tests are based on the comparison of two cumulative probability density functions, the sensitivity of Kuiper's test towards the boundaries of the distribution is higher. This test is, therefore, suited for studying the differences between lumiosity distributions at the faint end, and our reults support the notion that there is no truncation of the red sequence in high-redshift clusters, but that there is truncation in the field.

\textcolor{black}{The analysis of the luminosity distributions corroborates the results of the luminous-to-faint ratio, supporting the notion of the absence of truncation at $M_V > M_V^*$ in the HCS sample. Our results are also in agreement with \cite{Haines_2008}, who studied galaxies in the Sloan Digital Sky Survey (SDSS, \citealt{York_2000}) finding a deficit at the faint end in the field red sequence and no deficit in the cluster red sequence. A deficit in the field at the faint end of the red sequence is also observed in \cite{Bell_2004} and \cite{Brown_2007} at $z<1.0$. \cite{Mancone_2012} find no significant difference between the faint-end slopes of the IR luminosity functions in clusters at high ($1.0 < z < 1.4$) and low ($z < 0.4$) redshifts, although these author do not consider the red and blue galaxy populations separately.}

\textcolor{black}{In an analysis of the red sequence dwarf-to-giant ratio in clusters and in the field at $z < 1.0$, \cite{Gilbank_2008} find that this quantity, which is the inverse of the luminous-to-faint ratio, is at all redshifts lower in the field than in the clusters. This conclusion is in agreement with our results and with \cite{Haines_2008}.}

Our results suggest that the growth of the red sequence at low masses is accelerated in clusters of galaxies. The dense cluster environment favours mechanisms such as ram-pressure stripping, which has been shown to be the main driver of star formation quenching in low-mass galaxies in local clusters \citep{Gavazzi_2013}, although tidal interactions and galaxy harassment and strangulation can also lead to the accelerated consumption and/or the loss of cold gas (\citealt{Moore_1998}, \citealt{Bekki_2011}, \citealt{Larson_1980}), depleting galaxies of fuel for new star formation. Discriminating among all these mechanisms is beyond the scope of this paper, whose aim is the study of the observational properties of the red sequence in clusters at $z \sim 1$. However, it is important to stress that most clusters and proto-clusters at $z>1.5$ are not virialised and are composed of groups merging on to each other (\citealt{Hayashi_2012}, \citealt{Yuan_2014}). This suggests that the build-up of the red sequence may already take place in the groups ({\itshape{preprocessing}}, \citealt{Li_CNOC_2009}, \citealt{Taranu_2014}).

We note that most studies in the recent literature report the existence of a deficit of galaxies at the faint end of the red sequence in high-redshift clusters (\citealt{De_Lucia_2007}, \citealt{Gilbank_2008}, \citealt{Capozzi_2010}, \citealt{Bildfell_2012}, \citealt{Rudnick_2012}, \citealt{Fassbender_2014}), supporting the scenario that, as observed in low-density environments, low-mass galaxies are less efficient in quenching star formation. Our results do not rule out this scenario but rather support the notion that this property may be typical of low halo mass systems. Indeed, we observe that the red sequence of XMMXCS2215 is less populated at faint magnitudes (Figure 6 top-right panel), and the sample of EDiscCS clusters at $0.4<z<0.8$ used in \cite{De_Lucia_2007} was composed of systems which are mostly at $M_{DM}< 5 \times 10^{14} \mbox{} M_\odot$. The EDiscCS red sequence luminosity distributions in \cite{Rudnick_2009} all show a decreasing faint-end slope, similarly to what observed in our UltraVISTA passive red sequence sample and in agreement \textcolor{black}{with what is observed in} the stellar mass functions estimated in \cite{van_der_Burg_2013} for passive galaxies in the Gemini Cluster Astrophysics Spectroscopic Survey (GCLASS). This sample is composed of clusters of galaxies at redshift $0.8 < z < 1.3$ in the same halo mass range of EDisCS. 

While the disagreement between the results of this paper and those in the recent literature may be driven by the underlying physical properties of the sample, we stress that selection effects related to the construction of the sample may also introduce systematic errors. We have tested the number counts and $L/F$ using two different selection criteria for the red sequence and found that our conclusions remain unchanged. Nevertheless, \cite{Taylor_2015} showed that different selection techniques may affect the stellar mass distribution of the red sequence at $\log(M_*/M_\odot) < 9.5$. Although the HCS sample is not complete at such low masses (\citealt{Delaye_2014}), this effect should be taken into account in analysing deeper samples.

A direct comparison with the results of \cite{De_Lucia_2007}, \cite{Gilbank_2008}, \cite{Capozzi_2010}, and \cite{Bildfell_2012} on the evolution of $L/F$ is not possible with our data because we are limited to magnitudes $M_V < -19.5$ mag, while those authors were able to explore the red sequence down to $M_{V, Vega} = -18.2$ mag. The different magnitude ranges adopted for the definition of the bright and faint samples can influence the results of the measurements. One other source of disagreement is related to the selection of the sample, which may happen to be populated only by clusters that have a deficit of galaxies at the faint end of the red sequence. For example, \cite{Valentinuzzi_2011}, adopting the same magnitude ranges for the definition of the bright and faint samples, obtained a median $L/F$ consistent with the value found by \cite{De_Lucia_2007} at $z=0.57$. If one only compares these two samples, no truncation of the red sequence would be detected. After all, Figure 9 of \cite{De_Lucia_2007} shows that all measurements of $L/F$ in EDisCS are consistent within the errors, and that $L/F$ at $z=0.4$ is consistent with the values of $L/F$ found in their $z=0$ comparison samples. The uncertainties in $L/F$ are generally large because of the scatter between different clusters, as it can also be seen from \cite{Capozzi_2010} and \cite{Bildfell_2012} and, therefore, any conclusion regarding the existence of a truncation of the red sequence from such measurements should be treated with caution.

\textcolor{black}{Evidence for an accelerated suppression of star formation in clusters has come in the recent years from studies of the star-formation activity in distant clusters. There is evidence that the most distant clusters, at $z>1.5$, harbour actively star-forming galaxies in their cores (e.g.: \citealt{Hilton_2010}, \citealt{Tran_2010}, \citealt{Santos_2014}, \citealt{Santos_2015}). The works of \cite{Webb_2013}, \cite{Brodwin_2013}, and \cite{Alberts_2014} show that at high redshifts ($z>0.8$) star formation occurs at all radii in clusters, also in the cores. However, these author also show that the suppression of star formation occurs earlier in high-mass haloes than in low-mass haloes. In particular, \cite{Brodwin_2013} suggest that the main driver of the accelerated star formation quenching are galaxy mergers, which first trigger intense starbursts and then fuel AGN which suppress star formation in the merger remnant. \cite{Brodwin_2013} show that such a scenario agrees with the merger-driven mass growth invoked by \cite{Mancone_2010} to explain the evolution of the Schechter $M^*$ parameters in clusters at $z>1.3$. In order to occur mergers require low velocity dispersions, which are more typical in groups than in clusters. Therefore, the most massive clusters may have experienced this phase at earlier epochs compared to lower-mass systems, when they were still assembling from lower mass haloes.}

\textcolor{black}{The investigation of the build-up of the red sequence presented in this paper is complementary to the studies of \cite{Webb_2013}, \cite{Brodwin_2013}, and \cite{Alberts_2014}, and our conclusion support the notion of the accelerated star formation quenching in clusters of galaxies emerging from these works.}

Studies of the stellar populations of cluster galaxies at $z \sim 1$ suggest that low mass galaxies were quenched at later epochs with respect to more massive galaxies (\citealt{Demarco_2010}, \citealt{Muzzin_2012}, \citealt{Nantais_2013b}), hence supporting an evolutionary scenario in which stellar mass is the primary driver of the build-up of the red sequence, and the effect of the environment is to accelerate or act as an on-off switch for star formation quenching in galaxies. This scenario is in agreement with a truncation of the red sequence and does not exclude that a deficit can be detected at magnitudes fainter than those probed by current data. However, larger and more homogeneous samples are needed in order to carry out such analyses with higher accuracy and statistical significance.

\begin{figure*}
  \centering
	\subfloat[]{\label{fig:a}}\includegraphics[width=0.45\textwidth, trim=0.0cm 30.0cm 0.0cm 0.0cm, clip, page=3]{plot_HCS_RS_number_counts_comparisons}
        \hspace{0.01 cm}
        \subfloat[]{\label{fig:b}}\includegraphics[width=0.45\textwidth, trim=0cm 30.0cm 0.0cm 0.0cm, clip, page=1]{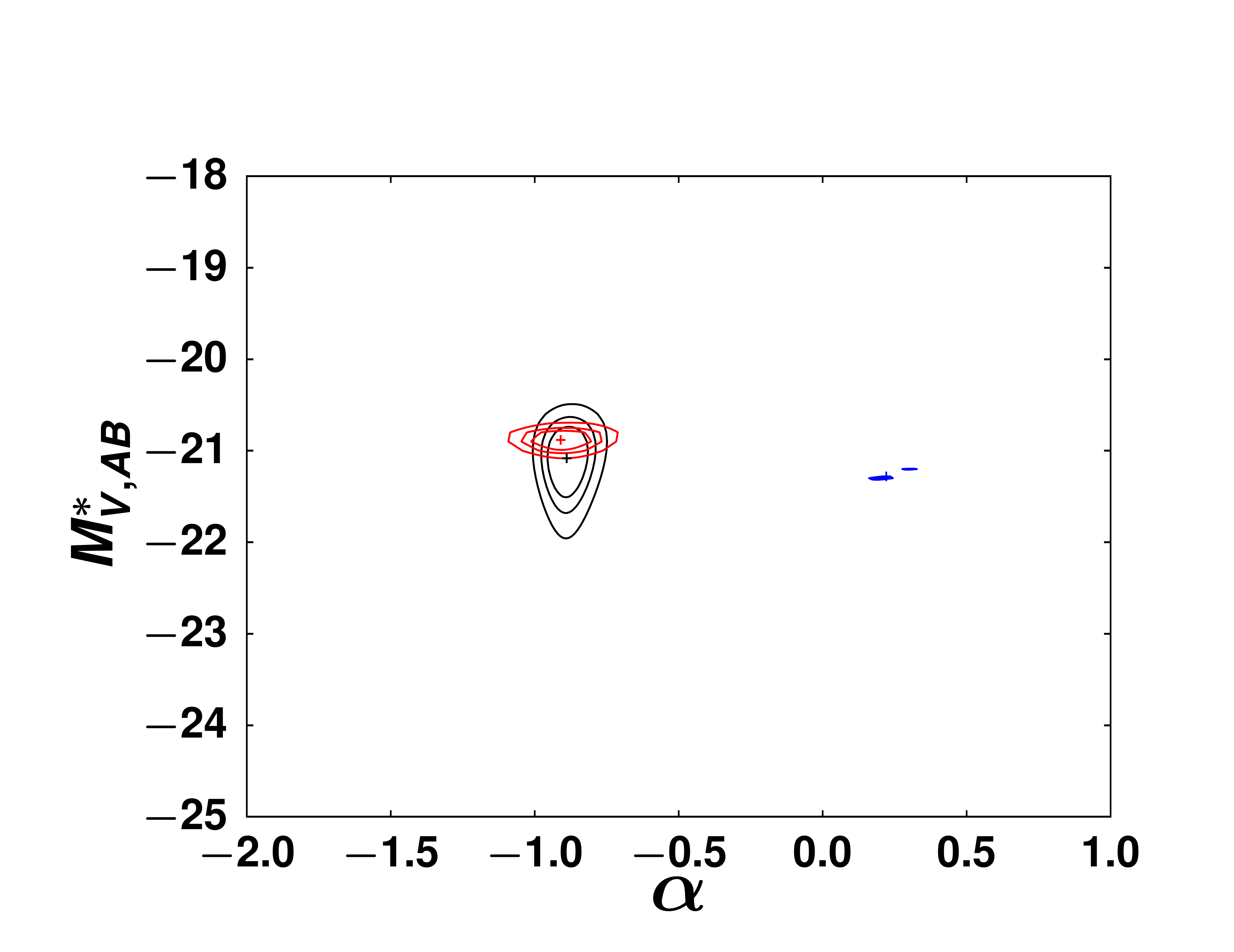}
	\caption{{\itshape{Left}}: Comparison between the red sequence number counts in clusters and in the field. Black filled circles represent the number counts of the entire HCS red sequence sample, while red diamonds are the red sequence number counts in the WINGS red sequence sample. The number counts of the HCS and WINGS composite samples were obtained as in \protect\cite{Garilli_1999} (see Section 4.5). Blue crosses are the red sequence number counts of passive red sequence galaxies in the COSMOS/UltraVISTA field at $0.8 < z_{phot} < 1.5$. The number counts in the WINGS and UltraVISTA samples are normalised to match the value of the HCS number counts at approximately the Schechter turn-over magnitude $M_{V}^*$. Solid lines are the best-fit Schechter curves (Equation 15) obtained for each sample. While no deficit of galaxies is observed in HCS with respect to WINGS, the UltraVISTA number counts decrease towards faint luminosities. This result suggests that the build-up of the red sequence is accelerated in clusters at low stellar masses. {\itshape{Right}}: confidence contours for the Schechter parameters $M_V^*$ and $\alpha$ derived for the HCS (black), WINGS (red), and UltraVISTA (blue) samples. The figure shows the 68\%, 90\%, and 99\% contours.}
        \label{fig:fig17}
\end{figure*}

\section{Summary and Conclusions}

We have presented a comprehensive analysis of the evolution of the red sequence in 9 galaxy clusters at $0.8<z<1.5$ drawn from the HAWK-I Cluster Survey (HCS, \citealt{Lidman_2013}). We studied the cluster red sequence focusing on two main aspects, the evolution of the red sequence zero-point, slope, and intrinsic scatter, and the build-up of the red sequence as a function of galaxy luminosity. Using the deep optical and NIR data and the spectra available for these clusters and presented in Section 2, we performed accurate PSF-matched photometry and estimated the parameters of the red sequence. The GOODS-N and GOODS-S fields observed at the same wavelengths of the HCS clusters were used as control fields to statistically estimate the contamination of the red sequence from field interloper galaxies. Our main conclusions are the following.

\textcolor{black}{We find no significant evolution in the red sequence slope and scatter, while the $(B-V)$ zero-point becomes redder towards lower redshifts, in agreement with the predictions of stellar population models. Our results suggest that the cluster red sequence was already assembled at $z=1.5$, and its component galaxies evolved passively until $z=0$.} This result is in agreement with all recent observational studies of the red sequence in clusters at redshifts $z<1.8$ but is at odds with the predictions of the hydrodynamical and N-body simulations of \cite{Romeo_2008} and \cite{Menci_2008}.

The luminous-to-faint ratio $L/F$ of red sequence galaxies does not evolve with redshift, in agreement with the works of \cite{Andreon_2008}, \cite{Lidman_2008}, \cite{Crawford_2009}, and \cite{De_Propris_2013} but in contrast with the conclusions of most authors in the recent literature (e.g.\ \citealt{De_Lucia_2007}, \citealt{Gilbank_2008}, \citealt{Capozzi_2010}, \citealt{Bildfell_2012}, \citealt{Lemaux_2012}, \citealt{Rudnick_2012}, \citealt{Fassbender_2014}). The latter find that $L/F$ increases with redshift, suggesting a truncation of the red sequence at low masses in high redshift clusters. 

The HCS red sequence luminosity distribution has a faint end slope consistent with that measured in the low-redshift WINGS clusters, supporting the trend of the luminous-to-faint ratio. However, we find that the luminosity distribution of passive red sequence galaxies in the COSMOS/UltraVISTA field at $0.8<z<1.5$ shows a statistically significant decrease at faint luminosities.

Our results suggest that the mass of the host dark matter halo plays an important role in setting the timescales for the build-up of the red sequence. In particular, the build-up of the red sequence appears accelerated in the most massive haloes at the lowest stellar masses as a result of the interactions between galaxies favoured by the dense cluster environment.

\newpage

\section*{Acknowledgments}

We thank the anonymous referee for the constructive feedback and comments provided during the revision of this manuscript. The results presented in this paper expand two chapters of P.C.'s PhD thesis. We thank Michael Brown, Casey Papovich, and Ray Sharples for their constructive feedback and suggestions. This work was performed on the gSTAR national facility at Swinburne University of Technology. gSTAR is funded by Swinburne and the Australian Government’s Education Investment Fund. We thank the Swinburne supercomputing and ITS teams for their support. We thank Maurizio Paolillo for providing clarifications about methods for the construction of composite luminosity distributions. The up-to-date redshift catalogue for the RDCS1252 field was provided by Julie Nantais. We would like to thank Bianca Maria Poggianti and Alessia Moretti for providing us with the latest versions of the WINGS catalogues. We also thank Boris H\"{a}u{\ss}ler for providing the latest up to date scripts for galaxy image simulations. We thank Alessio Romeo for providing the red sequence slopes and scatters from the most up-to-date cosmological simulations realised in his team. P.C. is the recipient of a Swinburne Chancellor Research Scholarship and an AAO PhD Scholarship. W.J.C. gratefully acknowledges the financial support of an Australian Research Council Discovery Project grant throughout the course of this work. R.D.\ gratefully acknowledges the support provided by the BASAL Center for Astrophysics and Associated Technologies (CATA), and by FONDECYT grant N. 1130528. L.F.B.\'s research is supported by proyecto FONDECYT 1120676, Proyecto Financiamiento Basal PFB06, and to the Millennium Science Initiative through grant IC120009, awarded to The Millennium Institute of Astrophysics, MAS. The data in this paper were based in part on observations obtained at the ESO Paranal Observatory (ESO programme 084.A-0214). This paper is partially based on observations obtained at the Gemini Observatory, which is operated by the Association of Universities for Research in Astronomy, Inc., under a cooperative agreement with the NSF on behalf of the Gemini partnership: the National Science Foundation (United States), the National Research Council (Canada), CONICYT (Chile), the Australian Research Council (Australia), Minist\'{e}rio da Ci\^{e}ncia, Tecnologia e Inova\c{c}\~{a}o (Brazil) and Ministerio de Ciencia, Tecnolog\'{i}a e Innovaci\'{o}n Productiva (Argentina).

\appendix

\section{Colour Images of the HAWK-I Clusters}

In this appendix we show colour images of the clusters in the HAWK-I Cluster Survey (HCS) obtained with the data used in this paper. In all the images North is up and East is to the left of the figures. The scale, in units of projected kpc at the redshift of the clusters, is indicated in each figure. Clusters are ordered by increasing redshift. Red sequence galaxies are highlighted by white circles, while yellow circles indicate spectroscopically confirmed cluster members. A description of the individual clusters in the HCS can be found in the Appendix of \cite{Delaye_2014}. All the colour images are shown in Figure A1, while the filter bands used in their production are shown in Table A1.


\begin{figure*}
\begin{tabular}{cc}
  \includegraphics[width=0.5\textwidth, trim=0.0cm 0.0cm 0.0cm 0.0cm, clip]{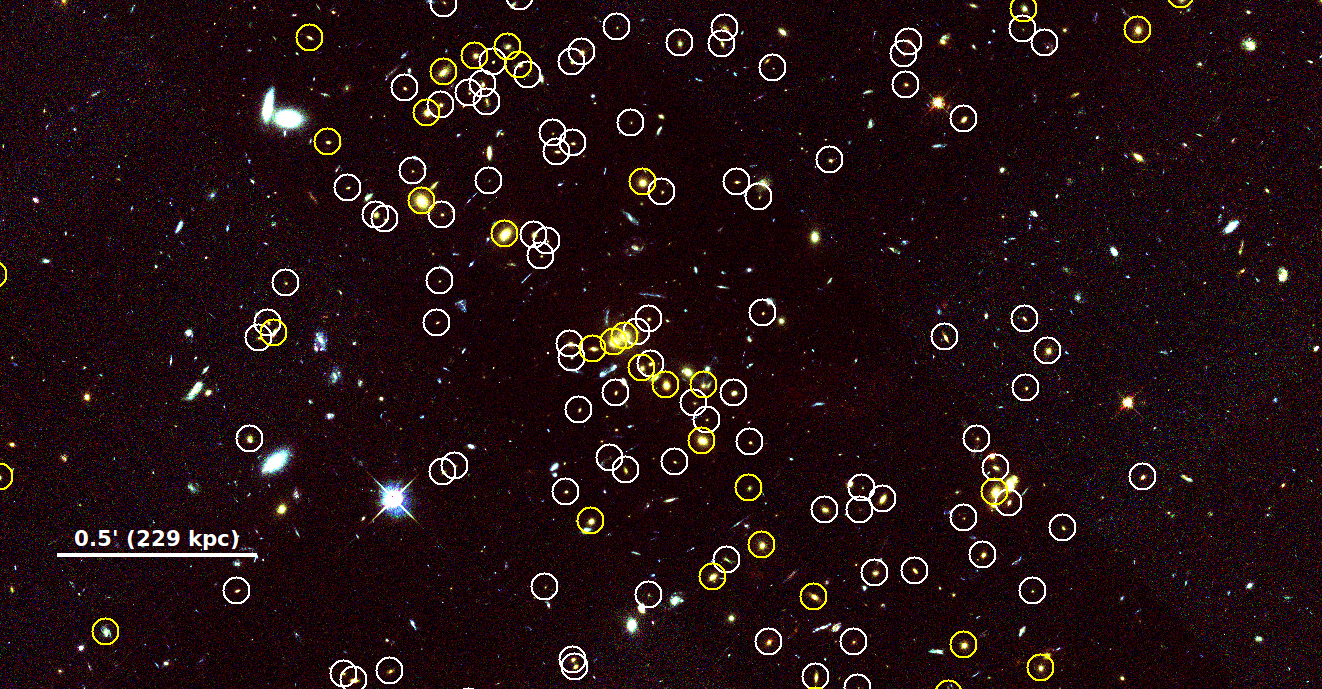} & \includegraphics[width=0.5\textwidth, trim=0.0cm 0.0cm 0.0cm 0.0cm, clip]{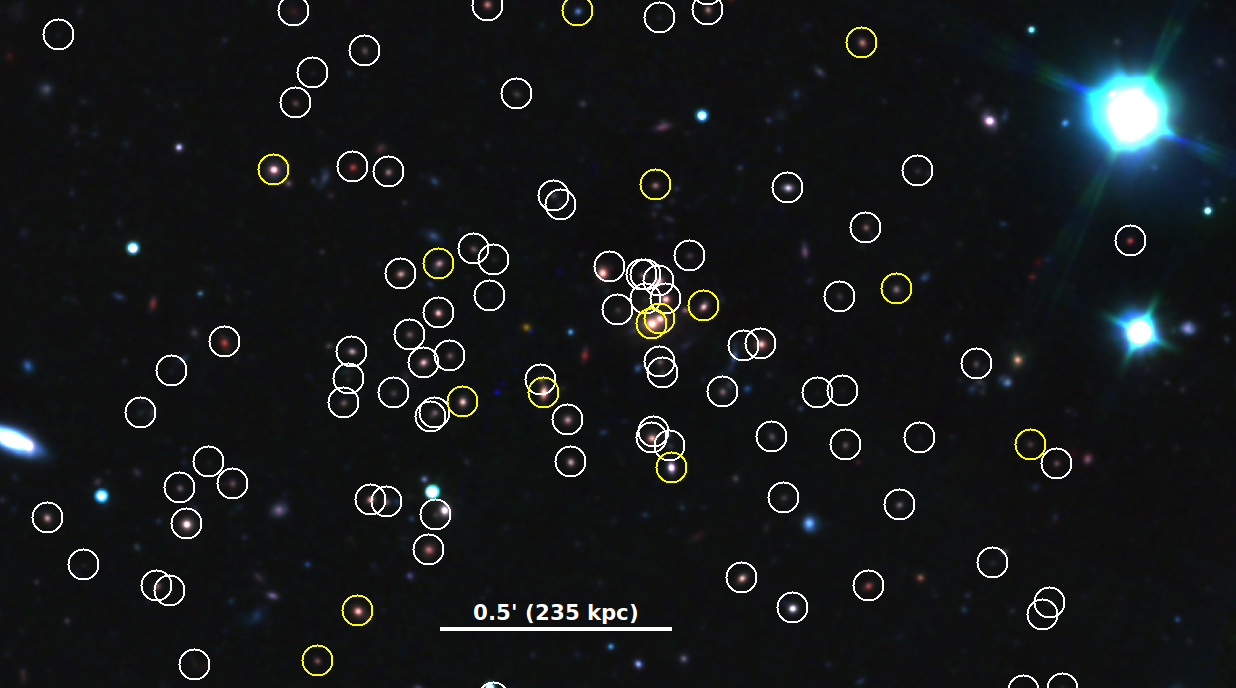} \\
         RX J0152.7-1357 (RXJ0152, $z=0.84$)       &    RCS 2319.8+0038 (RCS2319, $z=0.91$)          \\[6pt]
  \includegraphics[width=0.5\textwidth, trim=0.0cm 0.0cm 0.0cm 0.0cm, clip]{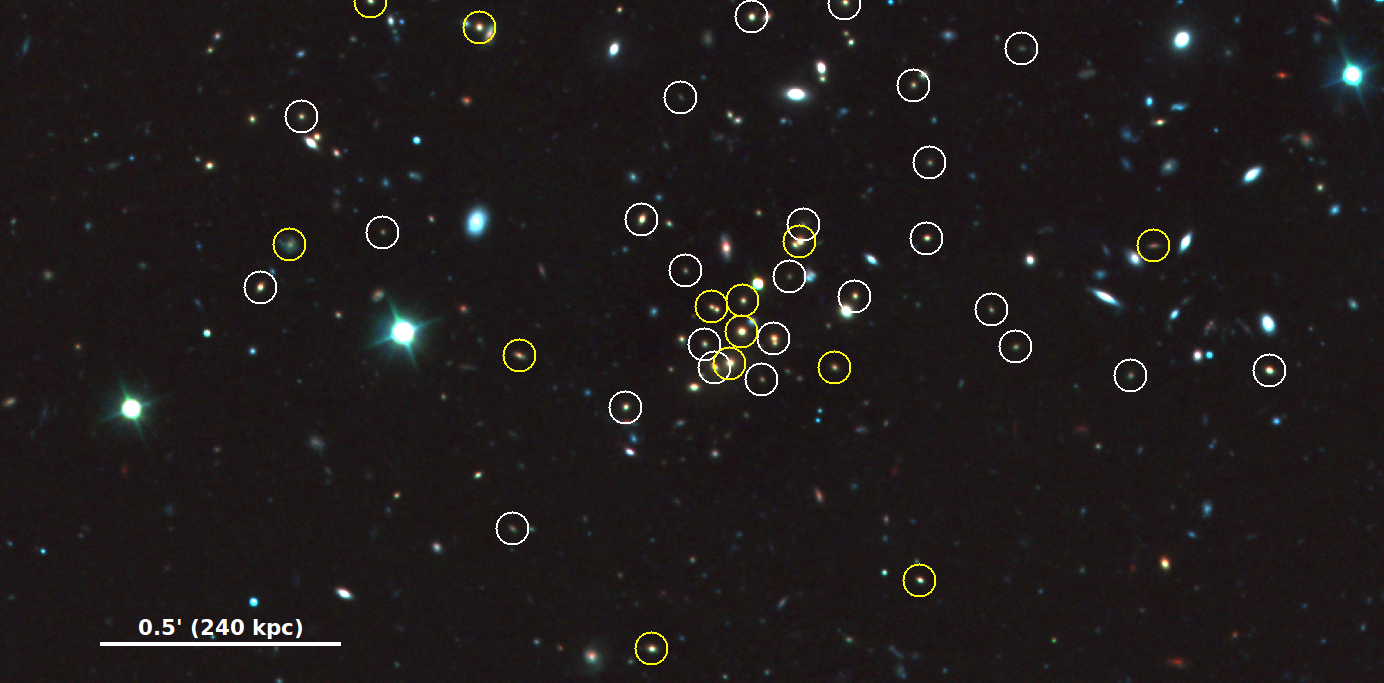} & \includegraphics[width=0.5\textwidth, trim=0.0cm 0.0cm 0.0cm 0.0cm, clip]{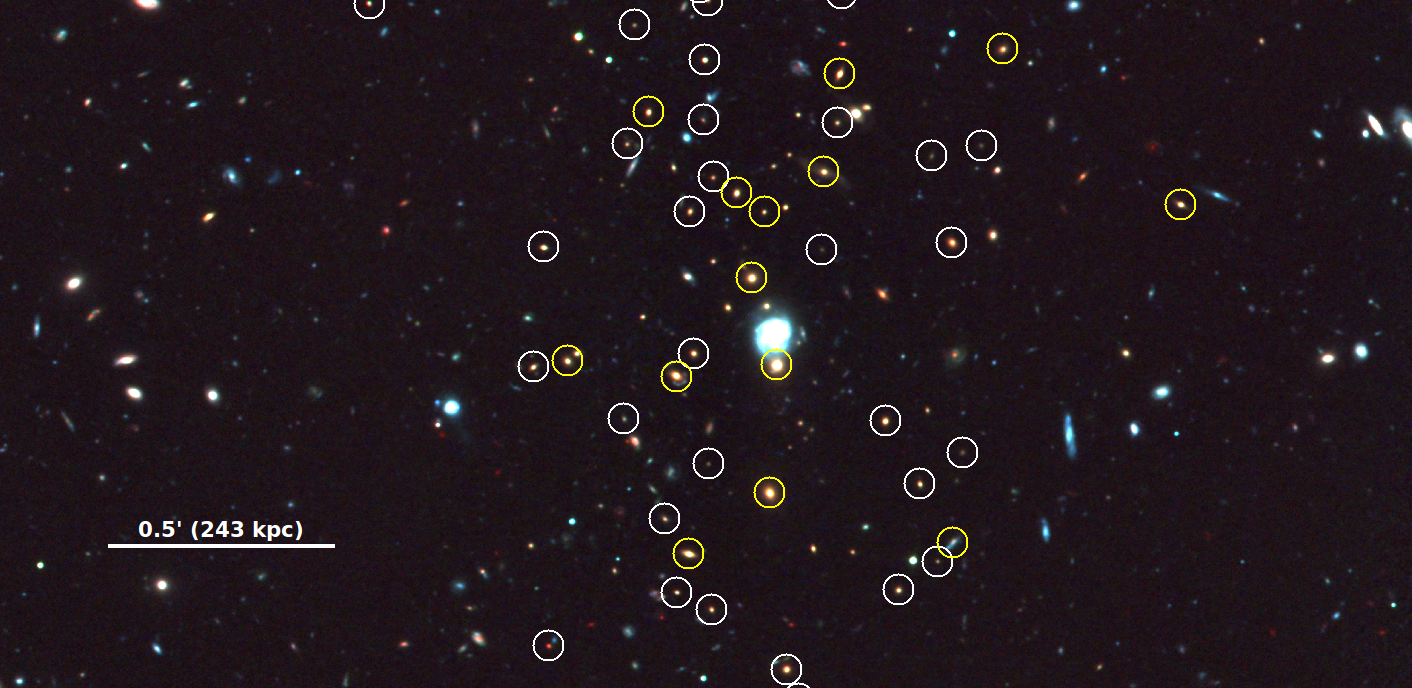} \\
          XMMU J1229+0151 (XMM1229, $z = 0.98$)    &    RCS 0220.9-0333 (RCS0220, $z=1.03$)             \\[6pt]
  \includegraphics[width=0.5\textwidth, trim=0.0cm 0.0cm 0.0cm 0.0cm, clip]{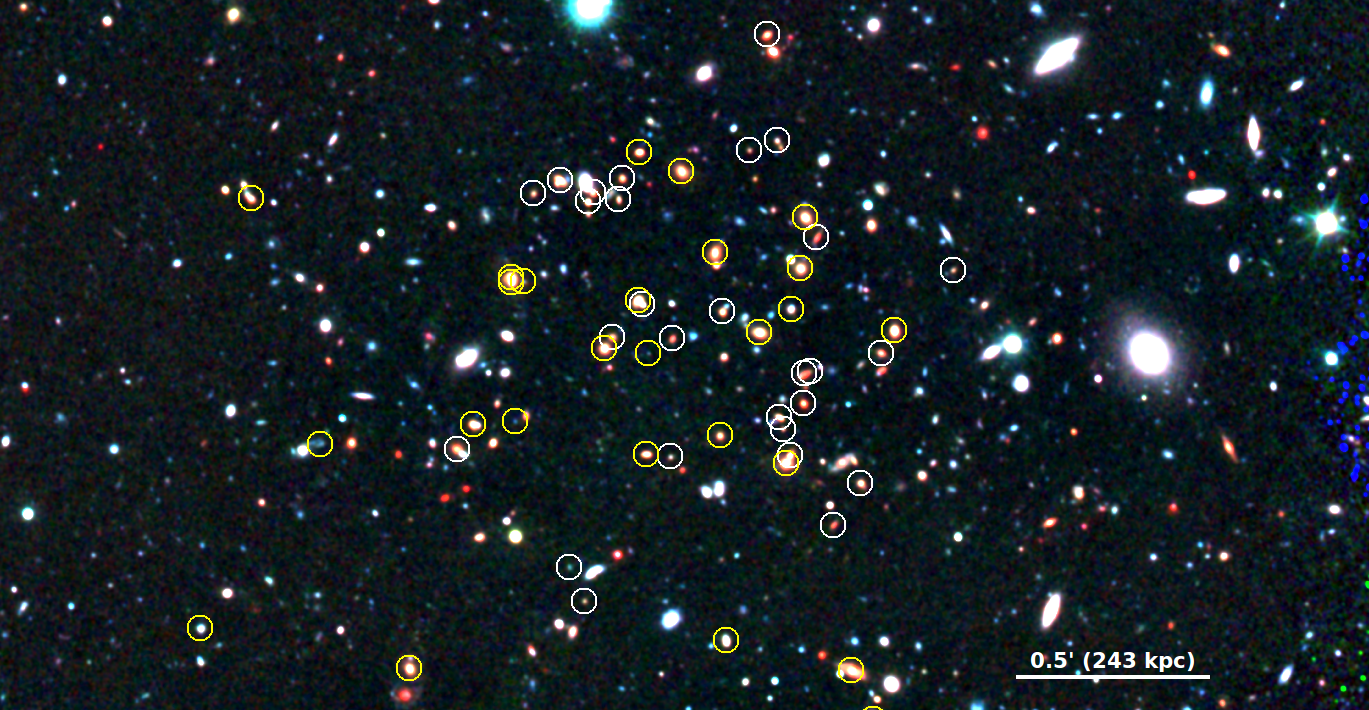} & \includegraphics[width=0.5\textwidth, trim=0.0cm 0.0cm 0.0cm 0.0cm, clip]{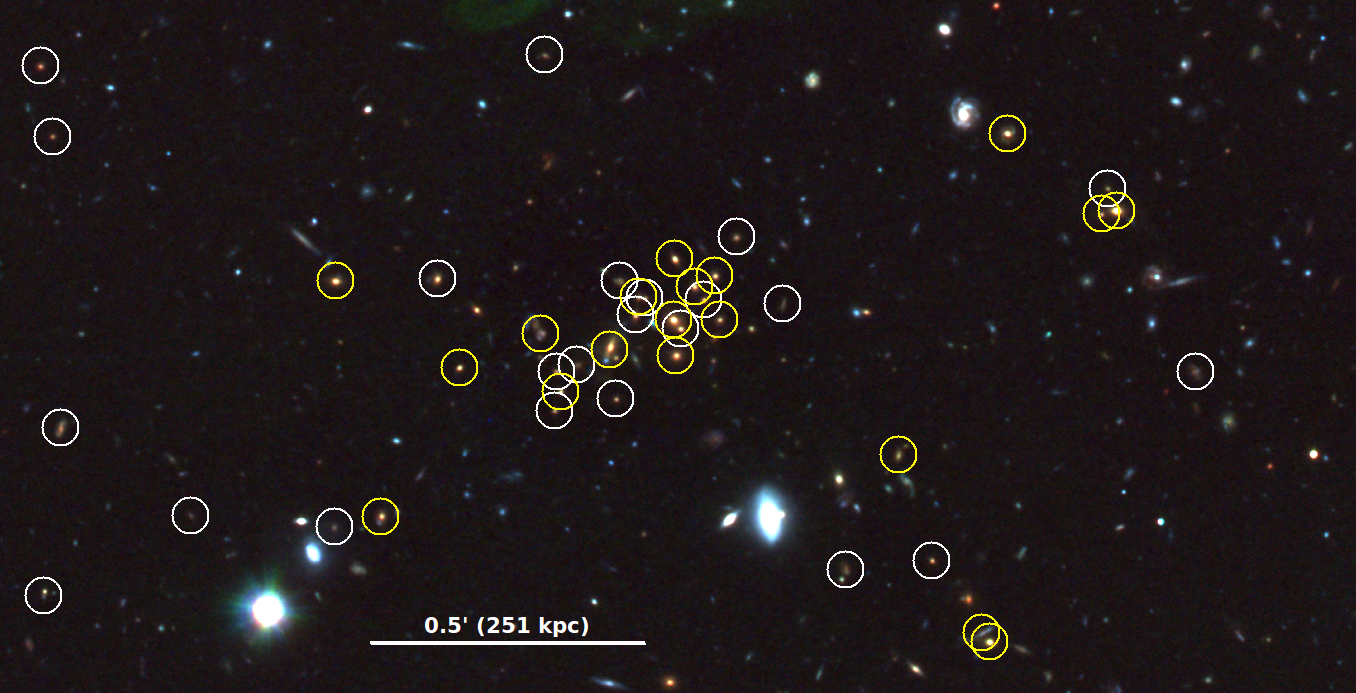} \\
          RCS 2345-3633 (RCS2345, $z=1.04$)        &    XMM J0223-0436 (XMMU0223, $z=1.22$)          \\[6pt]
  \includegraphics[width=0.5\textwidth, trim=0.0cm 0.0cm 0.0cm 0.0cm, clip]{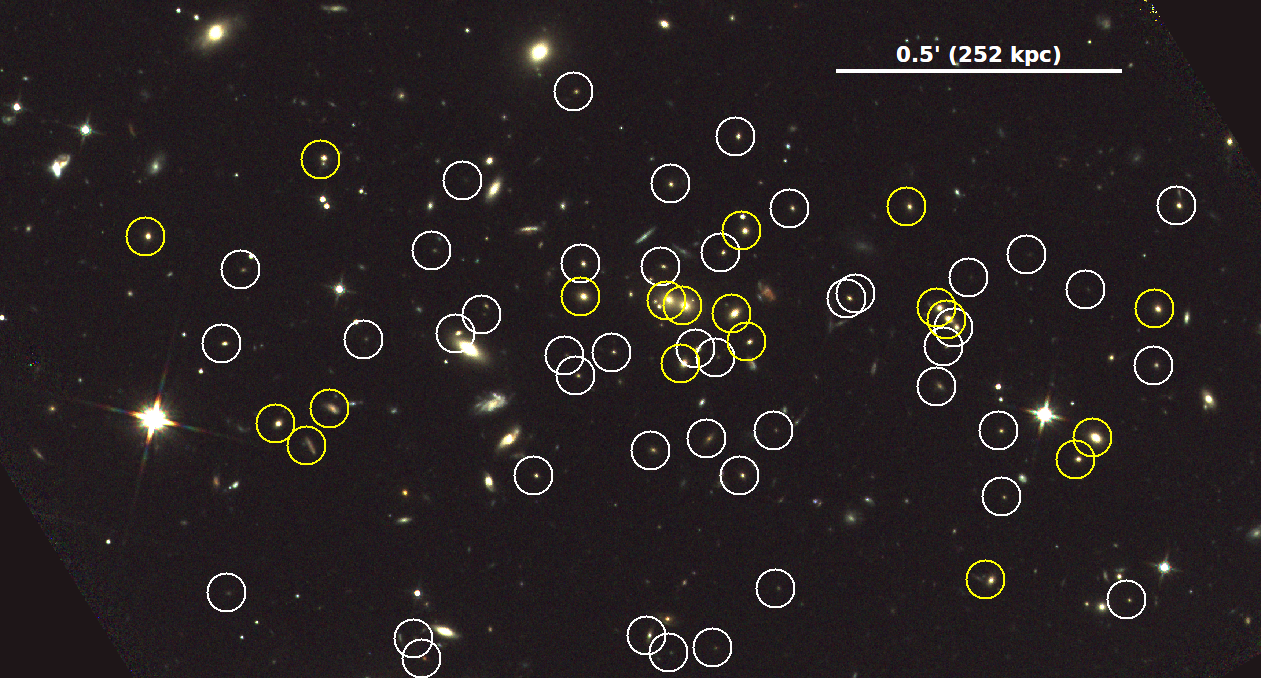} & \includegraphics[width=0.5\textwidth, trim=0.0cm 0.0cm 0.0cm 0.0cm, clip]{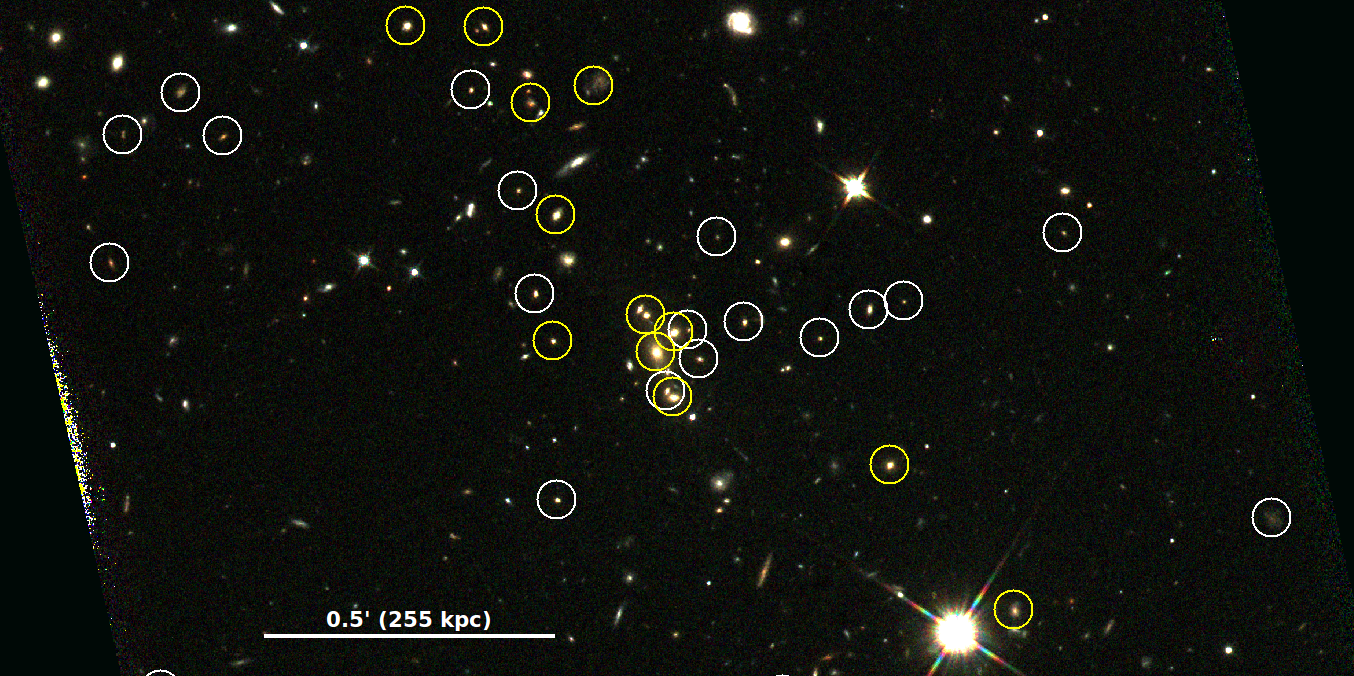} \\
          RDCS J1252.9-2927 (RDCS1252, $z=1.24$)   &    XMMU J2235.3-2557 (XMMU2235, $z=1.39$)          \\[6pt]
\end{tabular}
\caption{Colour images of the HCS clusters. The filter bands used in the production of the colour images are summarised in Table A1. In each figure North is at the top and East to the left. White circles highlight red sequence members while yellow circles highlight all the spectroscopically confirmed cluster members. The scale, in units of projected kpc at the cluster redshift is shown in each figure.}
\end{figure*}

\begin{figure*}
  \ContinuedFloat 
  \begin{tabular}{cc}
   \includegraphics[width=0.5\textwidth, trim=0.0cm 0.0cm 0.0cm 0.0cm, clip]{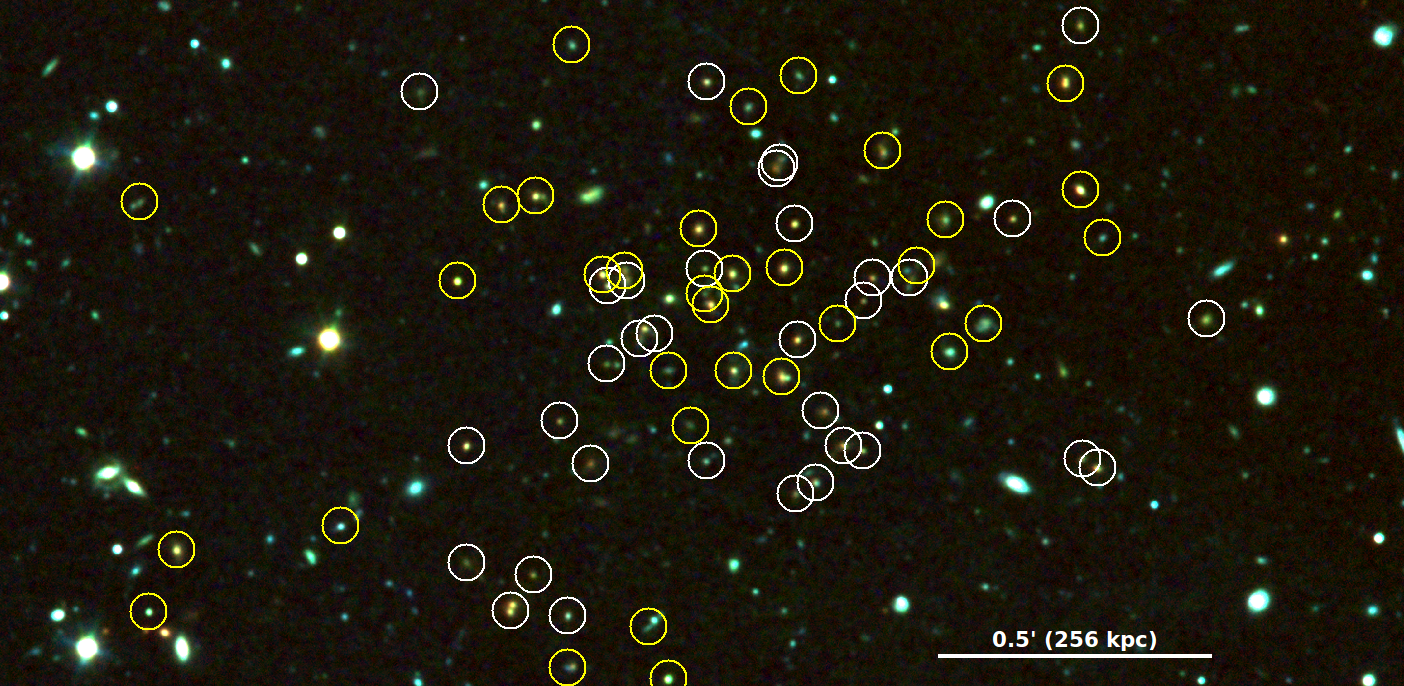} &  \\
        XMMXCS J2215-1738 (XMMXCS2215, $z=1.46$)      &              \\[6pt]
\end{tabular}
  \caption{Continued.}
\label{fig:fig3b}
\end{figure*}

\begin{table*}
  \caption{Filter bands used in the production of the HCS colour images shown in Figure A1. Filter names are ordered by increasing central wavelength from left to right. $Ks$ refers to the HAWK-I $Ks$ filter.}
   \begin{tabular}{|c|c|c|}
  \hline
     \multicolumn{1}{|c}{Cluster Name}  & \multicolumn{1}{c}{redshift ($z$)} & \multicolumn{1}{c}{Filter Bands} \\
      \hline
      \hline
      RX0152      & $0.84$ & $r_{625}$, $i_{775}$, $z_{850}$ \\
      RCS2319     & $0.91$ & $i_{775}$, $z_{850}$, $Ks$ \\
      XMM1229     & $0.98$ & $i_{775}$, $z_{850}$, $Ks$ \\
      RCS0220     & $1.03$ & $i_{775}$, $z_{850}$, $Ks$ \\
      RCS2345     & $1.04$ & $i_{775}$, $z_{850}$, $Ks$ \\
      XMMU0223    & $1.22$ & $i_{775}$, $z_{850}$, $J_{HAWK-I}$ \\
      RDCS1252    & $1.24$ & $Y_{105}$, $J_{125}$, $H_{160}$ \\
      XMMU2235    & $1.39$ & $Y_{105}$, $J_{125}$, $H_{160}$ \\
      XMMXCS2215  & $1.46$ & $i_{775}$, $z_{850}$, $J_{HAWK-I}$ \\
  \hline
  \end{tabular} 
\label{table13}
\end{table*}

\bibliographystyle{mn2e}
\small
\itemindent -0.48cm
\bibliography{biblio}

\bsp

\label{lastpage}

\end{document}